\documentclass{aa}  
\usepackage{graphicx}
\usepackage{txfonts}
\usepackage[]{natbib}
\usepackage{color}
\bibpunct{(}{)}{;}{a}{}{,}
\newcommand{\doce}{\mbox{$^{12}$CO}}

\newcommand{\trece}{\mbox{$^{13}$CO}}

\newcommand{\jsc}{\mbox{$J$=6$-$5}}

\newcommand{\jtd}{\mbox{$J$=3$-$2}}
\newcommand{\jdu}{\mbox{$J$=2$-$1}}

\newcommand{\kms}{\mbox{km\,s$^{-1}$}}

\newcommand{\ms}{\mbox{$M_{\mbox{\sun}}$}}
\newcommand{\ls}{\mbox{$L_{\mbox{\sun}}$}}
\newcommand{\my}{\mbox{$M_{\mbox{\sun}}$\,yr$^{-1}$}}

\newcommand{\lsim}{\raisebox{-.4ex}{$\stackrel{\sf <}{\scriptstyle\sf \sim}$}}
\newcommand{\gsim}{\raisebox{-.4ex}{$\stackrel{\sf >}{\scriptstyle\sf \sim}$}}

\newcommand{\secp}{\mbox{\rlap{.}$''$}}

\begin{document}

   \title{Structure and dynamics of the inner nebula around the
     symbiotic stellar system R Aqr}
   
   \author{V. Bujarrabal
          \inst{1}
  \and M.\  Agúndez\inst{2}
          \and 
          M.\ Gómez-Garrido\inst{1}
          \and Hyosun Kim\inst{4}
  \and M.\ Santander-Garc\'ia\inst{3}
          \and J.\ Alcolea\inst{3}
          \and A.\ Castro-Carrizo\inst{5}
  \and  J.\ Miko\l ajewska\inst{6}
   }

   \institute{             Observatorio Astron\'omico Nacional (OAN-IGN),
              Apartado 112, E-28803 Alcal\'a de Henares, Spain\\
              \email{v.bujarrabal@oan.es}
                          \and
  Instituto de F\'isica Fundamental, CSIC, C/ Serrano 123, 28006
  Madrid, Spain  
              \and
             Observatorio Astron\'omico Nacional (OAN-IGN),
             C/ Alfonso XII, 3, E-28014 Madrid, Spain
             \and
             Korea Astronomy and Space Science Institute, 776, Daedeokdae-ro,
 Yuseong-gu, Daejeon 34055, Republic of Korea
 \and 
 Institut de Radioastronomie Millim\'etrique, 300 rue de la Piscine,
 38406, Saint Martin d'H\`eres, France
 \and
  Nicolaus Copernicus Astronomical Center, Polish Academy
  of Sciences, ul.\ Bartycka 18, PL-00-716 Warsaw, Poland
   }

   \date{submitted April 2021, accepted April 2021}

   \abstract {} {We investigate the structure, dynamics, and chemistry
     of the molecule-rich nebula around the stellar symbiotic system R
     Aqr, which is significantly affected by the presence of a white
     dwarf (WD) companion. We study the effects of the strong dynamical
     interaction between the AGB wind and the WD and of
     photodissociation by the WD UV radiation on the circumstellar
     shells.}  {We obtained high-quality ALMA maps of the \doce\ \jdu,
     \jtd, and \jsc\ lines and of \trece\ \jtd. The maps were analyzed
     by means of a heuristic 3D model that is able to reproduce the
     observations. In order to interpret this description of the
     molecule-rich nebula, we performed sophisticated calculations of
     hydrodynamical interaction and photoinduced chemistry.}  {We find
     that the CO-emitting gas is distributed within a relatively small
     region \lsim\ 1\secp 5. Its structure consists of a central dense
     component plus strongly disrupted outer regions, which seem to be
     parts of spiral arms that are highly focused on the orbital
     plane. The structure and dynamics of these spiral arms are
     compatible with our hydrodynamical calculations. We argue that the
     observed nebula is the result of the dynamical interaction between
     the wind and the gravitational attraction of the WD. We also find
     that UV emission from the hot companion efficiently
     photodissociates molecules except in the densest and best-shielded
     regions, that is, in the close surroundings of the AGB star and
     some shreds of the spiral arms from which the detected lines
     come. We can offer a faithful description of the distribution of
     nebular gas in this prototypical source, which will be a useful
     template for studying material around other tight binary systems.}
             {}
  

   \maketitle
%

\section{Introduction}

Symbiotic stellar systems (SSs) are tight binary systems that typically
consist of an evolved Asymptotic Giant Branch star (the AGB primary,
very bright) and a white dwarf (WD, the secondary), in which
strong interactions are taking place, see, for instance, \cite{mikol12}. The
AGB wind is found to be strongly disrupted by the interaction, and an
active accretion disk is formed around the secondary WD, leading to
the ejection of very fast jets and nova-like phenomena.  The study of
SSs is essential for understanding all interacting binaries that
include evolved giants, in which the symbiotic-like activity is
usually much more elusive to observers, with implications for the
formation of bipolar planetary nebulae, the interaction of nova ejecta
with circumstellar material, the formation of chemically peculiar stars,
and the progenitors of type Ia supernovae.

The gravitational interaction between the wind from the primary and the
compact secondary is thought to be the dominant phenomenon that
explains the SS activity, and it is expected to deeply shape the
circumstellar wind nebula \citep[e.g.,][]{mohamedp12, valborro17,
  saladino18, kim19}. Hydrodynamical calculations show the efficiency
of a mass-transfer mode called wind Roche-lobe overflow (WRLOF), which
occurs when the wind velocity is still moderate when it reaches the
surroundings of the secondary, which significantly reinforces the
companion-wind interaction. After the interaction, the AGB-star wind is
predicted to be strongly focused on the binary orbital plane and to
form spiral-like arms or arcs.

R Aqr is the best-studied SS. The primary is a bright Mira-type
variable, and the companion is a WD. The nebula with size of 2$'$
consists of an equatorial structure that is elongated in the east-west
direction and a precessing jet (with a position angle, PA, ranging
between 10$^\circ$ and 45$^\circ$) powered by the accretion onto the
WD; see \cite{solfu85}, \cite{melnikov18}, and references therein. The
orbital parameters have recently been reanalyzed using new measurements
of the relative stellar positions and significantly more accurate
measurements of the AGB velocity \citep{bujetal18, alcmik21, 
gromik09}. The orbit shows a relatively long period,
$\sim$ 42 yr, and high eccentricity, with a typical stellar separation
equivalent to about 40 mas. At the time of our ALMA observations
(2017-2018), the WD was approaching a tight pericenter, which took
place by the end of 2019, and it was moving toward the observer at a
velocity of about 10 \kms. For our purposes, it is also important to
point out that the orbital plane is roughly perpendicular to the plane
of the sky, with its northern hemisphere pointing to us, and that the
orbital axis (projected on the plane of the sky) is placed at PA $\sim$
0-10 degrees. See \cite{alcmik21} and Sect.\ 4.1 for more details.

The properties of the primary are relatively well known. It is a very
bright M6.5-8.5 Mira-type variable, with a period of about 385 days.  A
relatively thick circumstellar envelope around it yields a moderate
extinction that varies with time in a not straightforward way that is
related to the geometry of the system, in particular, to the orbital
phase and the stream of gas that is ejected by the Mira primarily onto
the orbital plane \citep{gromik09}. Thus, measurements of
A$_\mathrm{v}$ range from 3.1~mag in 1977 (last periastron passage) to
0.5~mag only five years later \citep{brugel84}. Intermediate
values such as 1.9-2.0~mag were measured in between
\citep{wallerstein80,kaler81}, and an average of 0.7~mag (excluding
high-obscuration events) was found  in a period ranging from the mid-1970s to the
mid-2000s \citep{gromadzki09}. Typical extinction values of about 1-2
mag are therefore expected, but with significant variations with
time. The values of the extinction are particularly relevant for
estimating the effects of photodissociation (Sect.\ 4.3), but we recall
that the values toward the AGB and the WD can be different and that the
extinction also depends on the angle between the equator and the chosen
direction (the values given above only refer to the direction toward
the Earth, which is not in the equatorial plane).

The compact companion, whose effects on the nebula are expected to be
very strong, is more difficult to study because it is barely
observable. We discuss its properties in Sect.\ 4.3, in particular,
those that are relevant for our study of CO photodissociation.

Parallax measurements from SiO VLBI data indicate a distance of 218 pc
\citep[][]{min14}, although the distance from the GAIA parallax is 320
pc; the origin of this discrepancy is unknown. Both measurements are
subject to uncertainties because the stellar diameter and the
SiO-emitting region are much larger than the measured parallax; in
addition, both methods ignore the orbital movements of the mira
component. From our analysis of the orbital and stellar parameters, and
also accounting for the period-luminosity relations \citep{alcmik21},
we conclude that the most probable value is intermediate between them,
$\sim$ 265 pc.

Molecular emission is very rarely observed in SSs, probably due to
photodissociation by the UV emission of the WD and its surroundings
and strong disruption of the shells, except from regions very close to
the AGB. R Aqr is indeed the most intense SS  in  molecular lines, 
and its emissions have been relatively well studied
\citep[][]{bujetal10,bujetal18}. Although the CO lines in R Aqr are much
weaker than in standard AGB stars, they have been clearly detected and
mapped with ALMA.

\begin{figure*}
     \hspace{-.cm}
     \includegraphics[width=17.8cm]{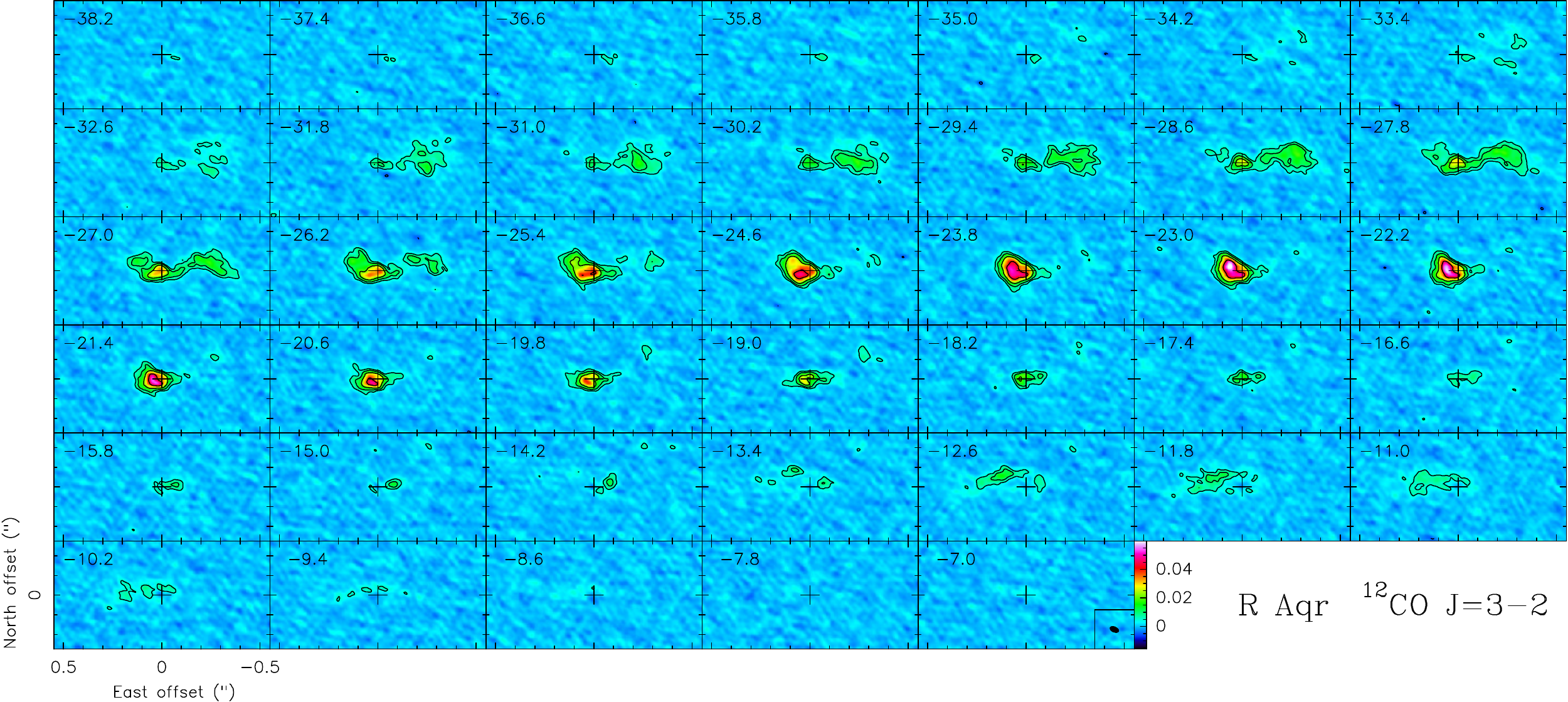}
      \caption{ALMA maps per velocity channel of \doce\ \jtd\ emission
        in R Aqr; see the {\em LSR} velocities in the upper left
        corners.  The center is the centroid of the continuum, whose
        image has been subtracted. The contours are logarithmic, with
        a jump of a factor 2 and a first level of $\pm$5 mJy/beam,
        equal to 6.2 times the rms and equivalent to 34.8 K (in
        Rayleigh-Jeans-equivalent brightness temperature units). The
        dashed contours represent negative values.  The HPBW is shown
        in the inset in the last panel.}
         \label{maps32}
\end{figure*}

\begin{figure*}
     \hspace{-.cm}
     \includegraphics[width=17.8cm]{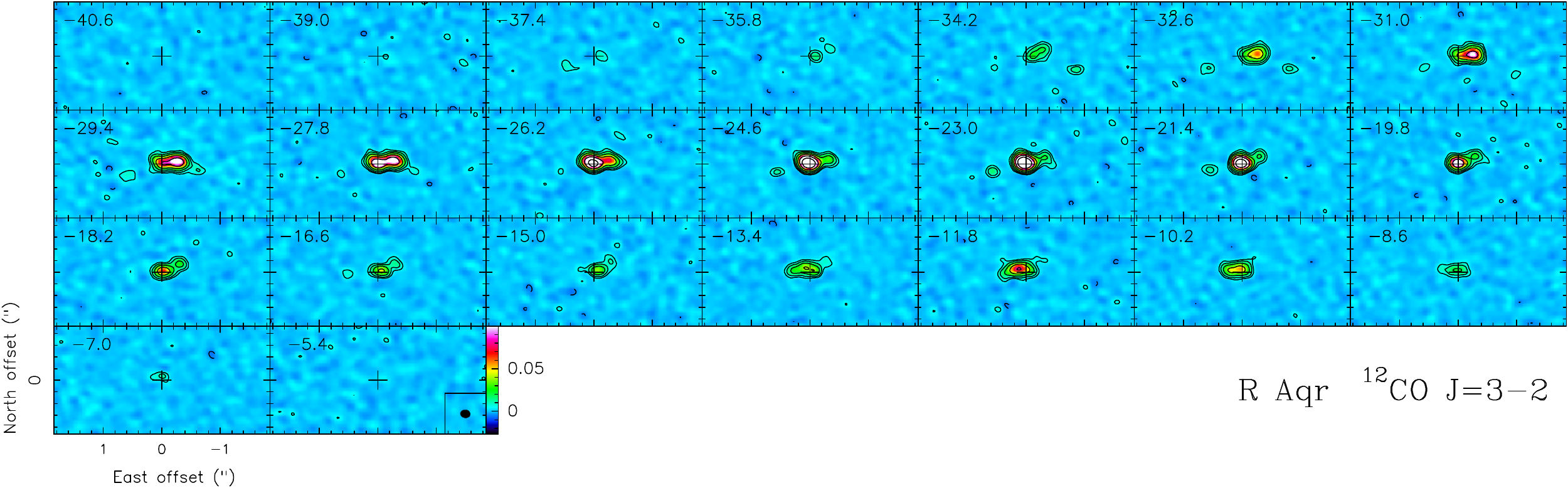}
      \caption{ALMA maps per velocity channel of \doce\ \jtd\ emission
        in R Aqr obtained with a more compact ALMA configuration; see
        the {\em LSR} velocities in the upper left corners.  The
        center is the centroid of the continuum, whose image has been
        subtracted. The contours are logarithmic, with a large jump of
        a factor 2 and a first level of $\pm$5 mJy/beam (equal to 3.4 times
        the rms and equivalent to 2.06 K). The dashed contours represent
        negative values.  The HPBW is shown in the inset in the last
        panel. The represented field is wider than in our other
        maps.}
         \label{ext32}
\end{figure*}

In our first ALMA observing run \citep{bujetal18, gomezg21}, we mapped
the surroundings of the star in continuum and line emission. We reached
suborbital scales in some cases. It is interesting to note that these
observations were obtained only three years after the
measurement of the stellar positions by \cite{schmid17}. In this paper,
we reanalyze some of our previous results and present new observations
of other lines of \doce, the only molecule showing well-resolved
emission with the ALMA angular resolutions (around 30 mas). These data
are analyzed by means of simple, heuristic modeling, whose main goal is
to describe the structure and kinematics that the inner nebula around R
Aqr must show to explain the observations. The various nebular
components were built by accounting for the results of very detailed
hydrodynamical calculations that were specifically performed for our case, and a
deep discussion of the chemistry, in particular, of the strong
photodissociation effects in this very complex environment.

\begin{figure*}
     \hspace{-.cm}
     \includegraphics[width=17.8cm]{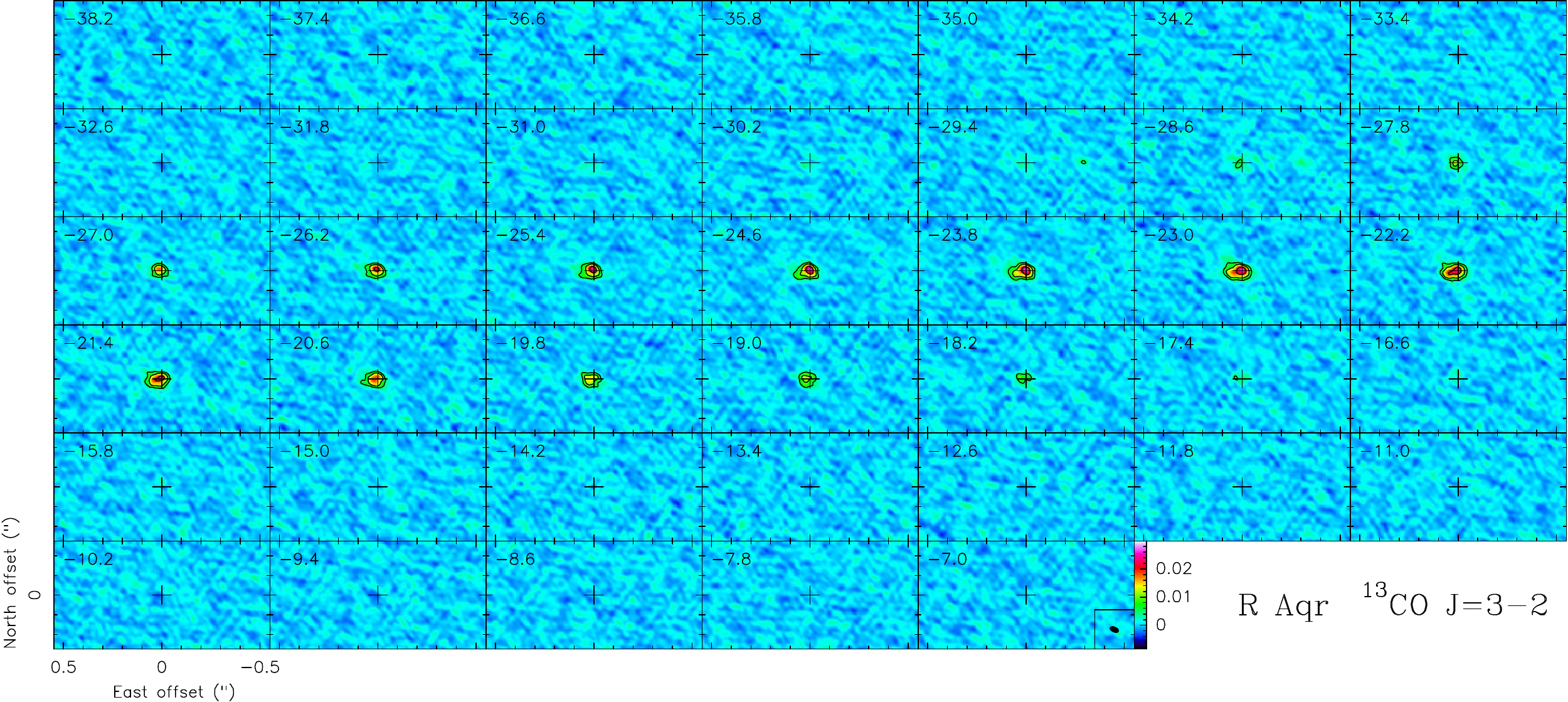}
      \caption{ALMA maps per velocity channel of \trece\ \jtd\ emission
        in R Aqr (only the more extended array configuration); see the {\em LSR}
        velocities in the upper left corners.  The center is the
        centroid of the continuum, whose image has been subtracted. The
        logarithmic contours and scales are the same as in
        Fig.\ \ref{maps32} (in this case, the first contour, 5
        mJy/beam, is equivalent to 5.6 times the rms and 35.2 K). The
        HPBW is shown in the inset in the last panel. }
         \label{maps13co32}
\end{figure*}

\section{New observations and reanalysis of previous data}

We present ALMA maps showing the extended emission of
the molecular lines \doce\ \jdu, \jtd, and \jsc, as well as maps of the
\jtd\ line of \trece. Resolutions between 20 and 200 milliarcsec (mas)
were attained. Maps of other lines and the continuum will be presented
in \cite{gomezg21}.

The high-resolution \doce\ \jtd\ observations were obtained with ALMA
in November 2017 and have been presented by \cite{bujetal18}. We
performed a new cleaning of the uv data using a new velocity channel
distribution to obtain comparable velocity brightness distributions for
all the lines presented here; a small spurious shift in the LSR
velocity (with negligible effect on the results) that was present in our
previous maps has been corrected. For more details on the data
acquisition and calibration, see \cite{bujetal18}. The maps are
presented in Fig.\ \ref{maps32}. The telescope half-power beam width
(HPBW) was 40$\times$30 mas, and the major axis at PA = 63$^\circ$. The
presented maps are centered on the continuum centroid, which for this
line is at ICRS coordinates R.A.: 23:43:49.4962, dec.: --15:17:04.720.

In the maps of this line, the continuum was subtracted from the
line maps (at the stage of uv data), to better show the emission in the
weak wings of the line. We proceeded the same way for the rest of the
maps presented in this paper, except for the \jsc\ map (see below). In
all our maps, except for the low-resolution \jtd\ maps, which are
mostly interesting to detect weak features, the data were resampled to
the same velocity resolution, 0.8 \kms. Because the movements of the
star are complex, with significant and uncertain proper movements and a
non-negligible orbit size, we decided to center all our maps on the
respective continuum centroids, which were measured independently for each
frequency and epoch. We recall that the dependence of the centroid
positions on the frequency and epoch of the observations is almost
unpredictable. All these effects are discussed in
\cite{gomezg21}. The centroid coordinates will be given in each case,
so that the meaning of the shifts can be assessed.

We also present maps of \doce\ \jtd\ obtained in September 2018 with a
more compact ALMA configuration, which therefore yield a poorer angular
resolution, but with a higher sensitivity (Fig.\ \ref{ext32}). The
calibration of the visibilities was performed using standard procedures
with the CASA software. The GILDAS software was used for the subsequent
cleaning, for which we used the Hogbom method and the Briggs weighting
scheme with a robust value of 1.  The HPBW in these observations is
0\secp 17$\times$0\secp 14 (PA = 83$^\circ$). These observations are
mainly useful to check the presence of more extended and weaker
emission. These high-sensitivity maps confirm the presence of weak
southern spots at a distance of between 0\secp 5 and 1$''$, which were
clearly detected and have been already depicted in maps by
\citet[][with a resolution of 0\secp 38$\times$0\secp
  33]{rams18}. The southern spots are about 30 times weaker than the
brightness peak, and therefore their detection in the high-resolution
maps is in principle not expected. These maps are also centered on the
corresponding continuum centroid, R.A.: 23:43:49.4994, dec.:
--15:17:04.765.

\begin{figure*}
     \hspace{-.cm}
     \includegraphics[width=17.8cm]{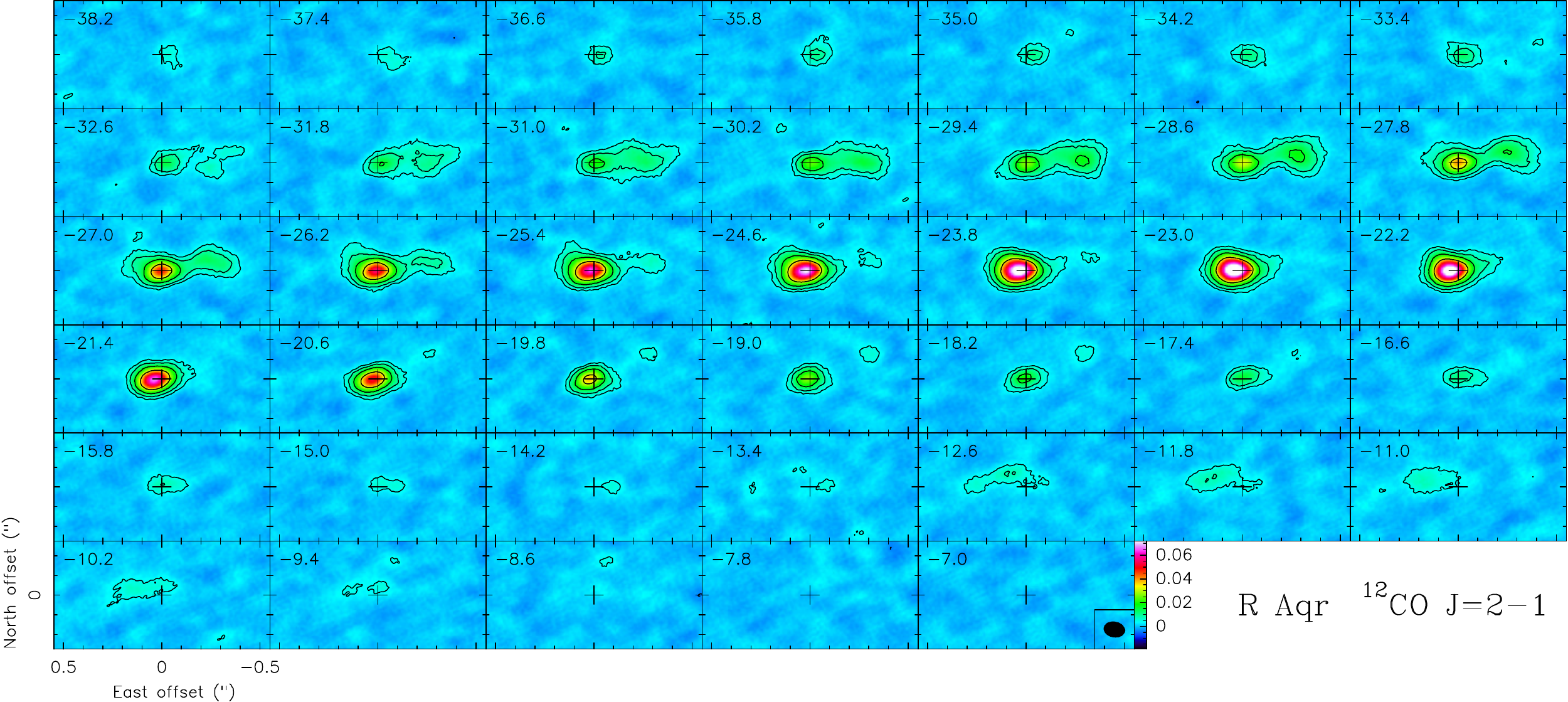}
      \caption{ALMA maps per velocity channel of \doce\ \jdu\ emission
        in R Aqr; see the {\em LSR} velocities in the upper left
        corners, similar to Fig.\ \ref{maps32}.  The contours are
        logarithmic, with a jump of a factor 2 and a first level of 4
        mJy/beam, equivalent to 4.9 rms and 11.33 K. The dashed contours
        represent negative values.  The HPBW is shown in the inset in the
        last panel.}
         \label{maps21}
\end{figure*}

\begin{figure*}
     \hspace{-.cm}
     \includegraphics[width=17.8cm]{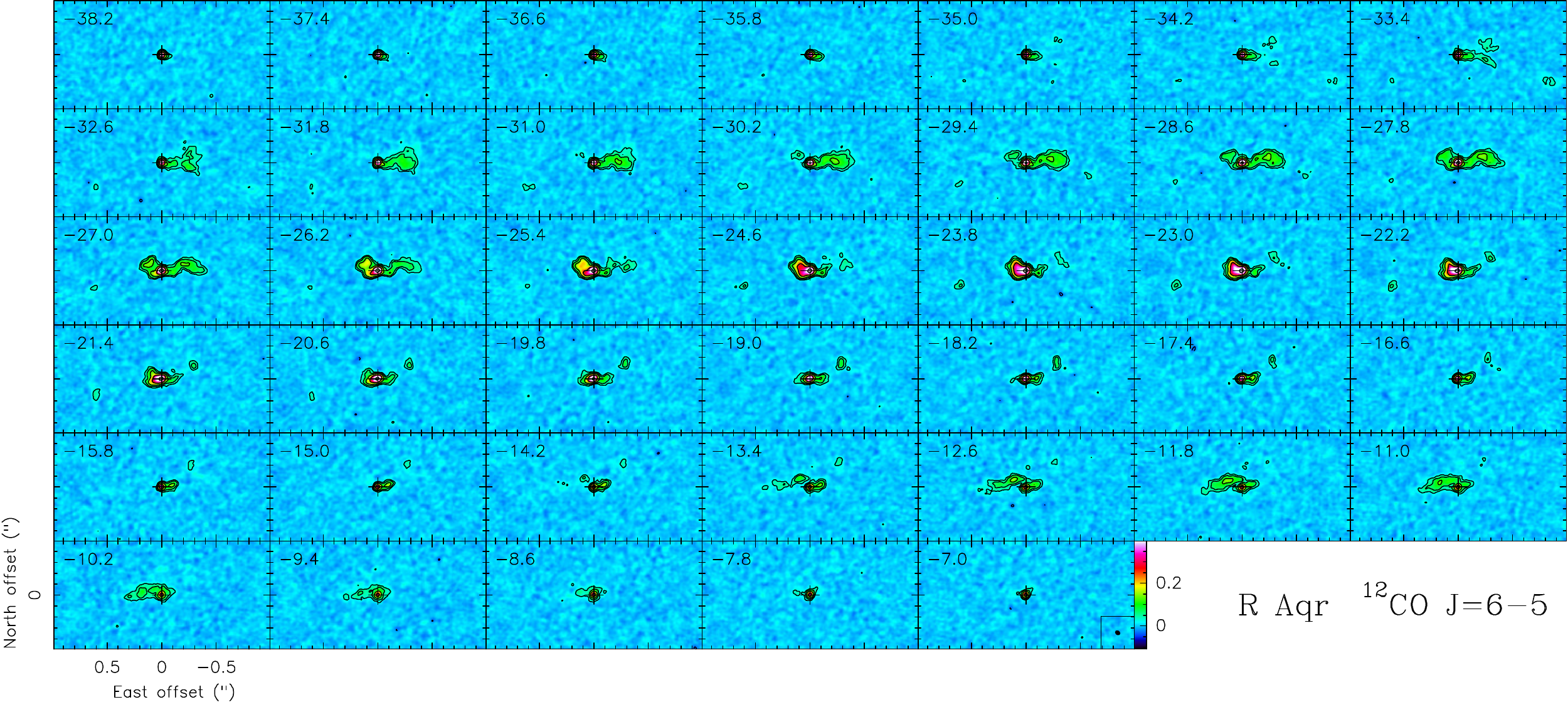}
      \caption{ALMA maps per velocity channel of \doce\ \jsc\ emission
        in R Aqr; see the {\em LSR} velocities in the upper left
        corners.  The contours are logarithmic, with a jump of a
        factor 2 and a first level of 50 mJy/beam, equivalent to 3.6 times
        the rms and 41.5 K. The dashed contours represent negative values.
        The HPBW is shown in the inset in the last panel. The velocity
        channels are the same as in Fig.\ \ref{maps32}, but the
        represented field is larger. The image center is the centroid
        of the continuum, whose image has not been subtracted
        (Sect.\ 2)}
         \label{maps65}
\end{figure*}

The comparison of the integrated flux-density profiles, F(Jy),
obtained from these low-resolution observations and the
wide-configuration observations presented before,
indicates a moderate loss of flux in the high-resolution maps of $\sim$
25\% at the central velocities (and \lsim\ 35\% at 10 \kms\ LSR).
There is no published single-dish profile to be compared with our
integrated flux, but this moderate flux loss is confirmed by our
\jdu\ maps, see below, whose integrated profile can be compared with
single-dish data and shows very similar fluxes. The brightness
distributions of all these (\jtd\ and \jdu) maps are completely similar
(taking the different resolutions into account and not considering the
weak detached spots).  We conclude that the flux lost in our
high-resolution \jtd\ maps is moderate, less than about one-third, and that it probably consists of  a contour at a level lower than
10\% of the peak values (in all the panels  of Fig.\ \ref{maps32}).

Maps of \trece\ \jtd\ observations were simultaneously obtained with the
high-resolution \doce\ map and were reduced in the same way and with very
similar instrument performances (Fig.\ \ref{maps13co32}). The
angular resolution is slightly worse due to the lower frequency,
50$\times$30 mas.

We also performed ALMA observations of \doce\ \jdu\ and \jsc. As
for \jtd\ observations, several other molecules were observed within
the simultaneous frequency bands. These weaker lines always show a much
less extended emission, the brightness being concentrated around the
AGB star. They will be discussed, together with the continuum
emission, in a forthcoming paper.

\doce\ \jdu\ was observed in June 2019 with an extended ALMA
configuration, leading to a beam size of 20$\times$20 mas with a major axis
at PA = 128$^\circ$.  The visibility calibration and map cleaning were
performed in the same way as described above.  Observations with a
more compact array were obtained in September 2019 and were reduced
in a similar way, yielding a resolution of 0\secp 20$\times$0\secp 15,
PA = 94$^\circ$.  Both observations were recentered on their
respective continuum centroids, which were found to be coincident
within the uncertainties (i.e., the difference was much smaller than
the beam of the low-resolution map). The result of merging both
maps is shown in Fig.\ \ref{maps21}, after centering on the merged
continuum centroid (R.A.: 23:43:49.4994, dec.: --15:17:04.765).
The resulting beam is 110$\times$80 mas with a major axis at PA =
--80$^\circ$.

The (angle-integrated) flux density of \doce\ \jdu\ shows a peak of
$\sim$ 0.2 Jy, with small differences between maps from more or less
compact configurations. As mentioned above, these values are compatible
with that from the single-dish profile in \cite{bujetal10},
and no significant flux seems to be lost in the \jdu\ ALMA mapping. It
is remarkable that the brightness distributions found in \doce\ \jdu\ and
\jtd\ (both compact and extended configurations) are very similar despite the 
different angular resolutions. We also discussed that this argues in
favor of just a moderate flux loss in \jtd\ and \jdu\ maps and a
negligible effect of the loss on the general distributions.

\doce\ \jsc\ was observed in August 2019, using an intermediate ALMA
configuration. The calibration and cleaning of this line were performed
as for previous observations, but we used a natural weighting
scheme to improve the map quality. The resulting beam is 50$\times$40
mas, PA = --49$^\circ$.  In this case, we did not subtract the
continuum because the line is highly opaque in the central velocities,
and subtraction in these panels results in a misleading relative minimum
(as background continuum emission is not added to the line emission
when the line opacity is high).  The resulting maps, centered again on the
continuum centroid (R.A.: 23:43:49.4996, dec.: --15:17:04.773), are
shown in Fig.\ \ref{maps65}.

\section{Observational results: Molecular line maps}

We present high-quality ALMA maps of \doce\ \jdu, \jtd, and \jsc\ and
of \trece\ \jtd, see Figs.\ \ref{maps32} to \ref{maps65} and
Sect.\ 2. Observations of other lines and the adjacent continuum are
presented in \cite{bujetal18} and \cite{gomezg21}.

In Fig.\ \ref{maps32} we show our high-resolution \jtd\ maps that were before
discussed in \cite{bujetal18}, after a reanalysis that included
rebinning of the velocity channels. Fig.\ \ref{ext32} presents maps
obtained with a more compact configuration, which yields a somewhat
poorer resolution but higher brightness sensitivity. The emission structure is
better described in Fig.\ \ref{maps32}, but the data in Fig.\ \ref{ext32}
confirm the detection of very weak clumps (tentatively detected in the
high-resolution maps) at offsets ($\Delta\alpha$,$\Delta\delta$)
$\sim$ (0.6,$-$0.2) and ($-$0.9,$-$0.2) arcsec. There is also a closer
component at approximately ($-$0.3,0.1), which is somewhat wider and
appears in a larger velocity range, between $-$23 and $-$12
\kms\ LSR. These clumps, particularly the last one, are also clearly
detected in \doce\ \jsc, see below.

Fig.\ \ref{maps21} shows observations of \doce\ \jdu, obtained
after merging observations with two different configurations.  The
brightness distribution is quite similar to that of \jtd, but weaker,
as expected because of the difference in frequency and
opacity. Precisely this difference in opacity is of particular
interest in our modeling (Sect.\ 4).

Finally, we show our \doce\ \jsc\ maps in Fig.\ \ref{maps65}. The weak
sparse clumps mentioned above are better detected in this line,
probably because of its higher opacity, which decreases the contrast
between high- and low-density regions. The higher opacity and
excitation requirements of this line are also useful in
modeling the CO emission. The represented field is larger in this
figure than for Fig.\ \ref{maps32} in order to show the weak clump
emission.

Some straightforward results can be directly extracted from our maps
of molecular line emission in R Aqr\,. We list them below.

\noindent $\bullet$ \doce\ \jtd, \jdu, and \jsc\ lines are resolved in
our ALMA maps. They occupy a region of about 0.7$\times$0.2 arcsec
(elongated more or less in the east-west direction). A few sparse, very
weak spots are detected at distances between 0\secp 5 and
1$''$. Despite the high sensitivity of our ALMA maps (we reach
very high signal-to-noise ratios, S/N $>$ 100 in most maps), these
clumps are barely detected.

\noindent $\bullet$ Both single-dish \citep{bujetal10} and
interferometric data indicate relatively weak total (angle-integrated)
fluxes, for instance, $F_{\rm peak}$(\jdu) $\sim$ 0.2 Jy. The
single-dish and angle-integrated ALMA profiles are very similar
(Sect.\ 2). The \doce\ line intensities are far lower and the extents
are far smaller than for standard AGB stars. For the same mass-loss
rate and distance ($>$ 10$^{-7}$ \my, $\sim$ 265 pc), we would expect
\doce\ \jdu\ peak fluxes $>$ 10 Jy and extents $>$ 10$''$
\citep[e.g.,][]{castroc10}. The CO lines in standard AGB sources are
about a hundred times more intense and extended than in R Aqr.

\noindent $\bullet$ Except for the \doce\ lines, the emission of all other
molecules in R Aqr is very compact \citep{bujetal18, gomezg21} and still much
weaker. Most lines occupy less that 0\secp 1 and even \trece\ lines are
more compact than 0\secp 2. In standard AGB envelopes, molecules other
than \doce\ show sizeable extents $>$ 1$''$ \citep[see,
  e.g.,][]{agundez17, verbena19}. The extent of \trece\ is often almost
comparable to that of \doce\ in AGBs because the lower degree of
selfshielding of \trece\ to the interstellar UV is partially compensated for by
fractionation reactions \citep[][etc]{cernicharo15, mamon87}.

\noindent $\bullet$
The \doce\ line maps show a central condensation, plus some plumes
that mostly extend in the east-west direction. The condensation is 
0\secp1-0\secp 2 wide. It is centered on the expected position of the AGB star
and slightly exceeds the full orbital region (the typical angular
separation between the two stars is up to $\sim$ 50 mas, see Sect.\ 1).

\noindent $\bullet$
In all three \doce\ lines, the extended plumes appear to be part of a
spiral structure that is mostly confined to the orbital plane and is
seen almost edge-on \citep[as was mentioned by][ this is more widely
  discussed in Sect.\ 4]{bujetal18}. The complex velocity structure
suggests a complex dynamics that probably includes both tangential and
radial expansion movements.

\noindent $\bullet$ The different observations were obtained at
different epochs, and because we were close to a tight periastron (that
took place at the end of 2019), fast variations in the relative stellar
positions are expected. In any case, all \doce\ maps show quite
comparable distributions.

\noindent $\bullet$ The fact that the \jsc\ line is the most
intense confirms that the lines come from high-excitation regions, with
temperatures that exceed its excitation energy ($E_{J=6}$ = 115 K), and
densities that are high enough to significantly populate these levels despite
its high A-coefficient (2 10$^{-5}$ s$^{-1}$), $n$ \gsim\ 10$^6$
cm$^{-3}$. The brightness peak in K-units, $\sim$ 550 K, implies gas
temperatures higher than this value in the central regions. In addition,
because the typical gas temperatures in the inner circumstellar envelopes are
usually not very high, about 1000 K at most, the opacity in this line
must be \gsim\ 1 in the central regions and somewhat lower elsewhere.

\noindent $\bullet$
In general, the dependence of the brightness level on the line
strength shows that the lines are not opaque: \jdu\ reaches a peak
brightness of 250 K, while \jtd\ is intermediate between
\jdu\ and \jsc. We conclude that optical depths must be moderate,
depending on the line and analyzed component, with opacities
between lower than 0.25 and $\sim$ 2. Our modeling (Sect.\ 4) confirms
this result.

\noindent $\bullet$ Our observations of
\trece\ \jtd\ (Fig.\ \ref{maps13co32}) confirm a significant opacity of
the \doce\ line in the central regions because the \doce/\trece\ line
ratio is low ($\sim$ 3), but not in the outer regions, in which practically no
\trece\ emission is detected. In any case, the abundance ratio must be low
($\sim$ 10-15), in view of the expected values of the opacities and the
moderate line intensity ratio. We also note that the maps, with
slightly resolved emission from this central clump, show a very simple
structure that is not compatible with a significant rotation, but suggest
simple expansion at moderate velocities \lsim\ 10 \kms.

\noindent $\bullet$ It is important to realize the strong difference
between the brightness distributions observed in R Aqr and those
typical of standard AGB stars \citep[see,
  e.g.,][]{castroc10,cernicharo15,guelin18,homan18, decin20}, which were obtained
using similar observational techniques. In most cases, the
circumstellar envelopes seem to be composed of more or less complete
spiral arms or arcs, which are probably the result of interaction with a low-mass
companion, but the overall structure is spherical and the kinematics is
roughly isotropic expansion. Only in some objects, such as the semiregular
variables R Dor and X Her, does the circumstellar envelope show an
hourglass-like structure with a clear symmetry axis. These peculiar
objects appear in some way to be intermediate between the usual roughly
spherical shells and the nebula around R Aqr.

\section{Modeling the ALMA maps of R Aqr}

As described in Sect.\ 3 and \cite{bujetal18}, a first inspection of the
ALMA maps strongly suggests that the CO line emission comes from spiral
arcs that are strongly focused on an equatorial plane and are seen roughly
edge-on. This might be compatible with results from
hydrodynamical calculations and the expected geometry of the binary
orbit and circumstellar shells in R Aqr (Sect.\ 1).  The amount of
information included in our high-quality maps allows a more detailed
comparison with predictions from the hydrodynamics of the system. For
this purpose, we performed an heuristic modeling of the data in
order to describe the main properties of the CO-rich detected nebula
(see Sect.\ 4.1), as well as hydrodynamical models of the gravitational
interaction between the WD companion and the AGB wind for the case of
the R Aqr system (Sect.\ 4.2).

From our first attempts to model the ALMA observations, we readily
found that the predictions from hydrodynamical models by themselves are
not able to provide an adequate density distribution, as they are in
general too extended to explain the small emitting region and the very
weak total CO emission. The very weak single-dish CO lines already
require a very limited emitting region, see Sect.\ 3 and
\citet{bujetal10}, instead of spiral arms occupying extended regions
(more than 10$''$ wide) that are globally similar to a standard AGB
circumstellar envelope. In addition, the ALMA maps only show a clearly
defined and limited group of spiral arcs, just some threads of the
inner spirals. Our simplified descriptive model of the CO-emitting
region (Sect.\ 4.1), based on the general properties of the
hydrodynamical predictions, is devoted to describing the maps in terms
of the emission of some inner arcs, as well as to defining the
properties that they must satisfy to reproduce the observations. We
further conducted new hydrodynamical calculations for the case of R Aqr
(Sect.\ 4.2), including in particular the effects of its significantly
eccentric orbit. The comparison between the descriptive and
hydrodynamical models will allow us to improve our understanding of the
effects of gravitational interaction between the compact companion and
the inner circumstellar shells.

Finally, we also try to understand the molecule photodissociation
due to the strong UV field of the white dwarf star. We show below
(Sect.\ 4.3) that photodissociation can be strongly selective and might
explain that only some pieces of the spiral structure show molecular
emission. We stress the extremely difficult treatment of the
chemistry in our source, which has a very complex structure and dynamics and
a significant variation with time (because the orbital period is
comparable to the abundance evolution times in many cases).

The goal of this modeling effort is to allow a sensible comparison
between the observations and the hydrodynamical simulations of the
strong gravitational attraction effects of the WD on the AGB
wind. Because the observations just probe the CO-rich regions, which
are probably affected by photodissociation, the chemistry induced by
the UV radiation of the WD must be also accounted for. In any case, the
results of our heuristic model, as they are basically a description of
the emitting region responsible for the maps, are actually closer to
the CO-rich gas distribution than any result from hydrodynamical and
chemical calculations, which are quite a-prioristic and mainly useful
for understanding the origin and meaning of the detected features.

\subsection{Descriptive, heuristic  modeling of the CO line emission from
  R Aqr}

\subsubsection{Main model characteristics and nebular components}
  
As mentioned, direct calculations of CO emission using predictions
directly from hydrodynamical models cannot explain some obvious
properties of the CO emission. In addition, a long iteration of
hydrodynamical calculations plus CO line emission simulations plus a
comparison with maps plus hydrodynamical calculations, etc., is not
feasible because hydrodynamical modeling is delicate and
time-consuming. Accordingly, we developed a descriptive model of the
CO-emitting region that accounts for the general properties of the
hydrodynamical predictions and the expected photodissociation
effects. This model describes the properties that the CO-rich gas
distribution and kinematics must show to reproduce the
observations. These properties are finally discussed based on
theoretical considerations.

The code we used for our heuristic model allows general 3D
distributions of the physical conditions and kinematics, whose
descriptions are quite free. Gas velocities can show both radial
(expansion or infall, though we favor expansion movements) and
rotational (i.e., tangential) components.  The contrast between
equatorial and adjacent regions could be very sharp and high, the
spiral arcs can be more or less complete, etc.  The only basic
  assumptions or ``axioms'' we adopt in our modeling are that there is
  a symmetry plane (the equator), which is the same as that of the
  orbit, and that the gravitational interaction can induce in the gas a
  tangential velocity in the direction of the movement of the
  companion, both expected from theoretical descriptions of
interaction in SSs, as summarized in Sect.\ 1. We also take previous
empirical information on the orbit (Sect.\ 1) into account, notably
that the axis of the orbit is almost in the plane of the sky, that its
north pole is slightly tilted to the east and pointing to us, and that
the secondary was moving toward the observer during the observing runs.

\begin{figure}
     \begin{center}
       \includegraphics[width=9cm]{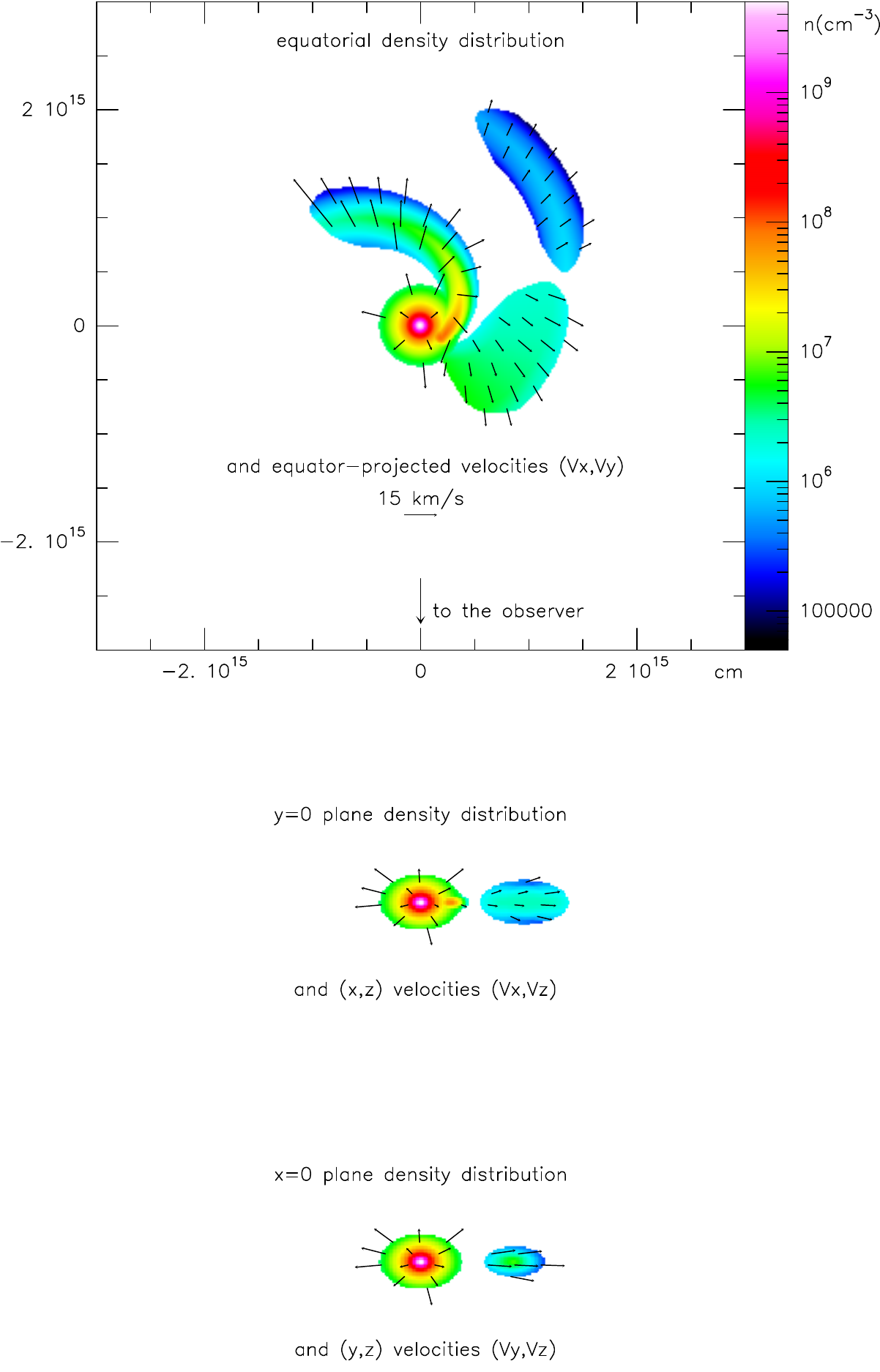}
       \vspace{.1cm}
     \caption{{\it top:} Density distribution of the CO-rich gas in the
       plane of the equator ($z$=0 plane in our coordinates) from our
       best-fit model. Equatorial velocities are also represented. {\it
         middle:} Density distribution in the $y$=0 plane and velocities
       projected in that plane. {\it bottom:} Same, but for the $x$=0
       plane. All representations share the same units and scales.}
         \label{mod}
     \end{center}
\end{figure}

\begin{figure*}
     \hspace{-.cm}
     \includegraphics[width=17.8cm]{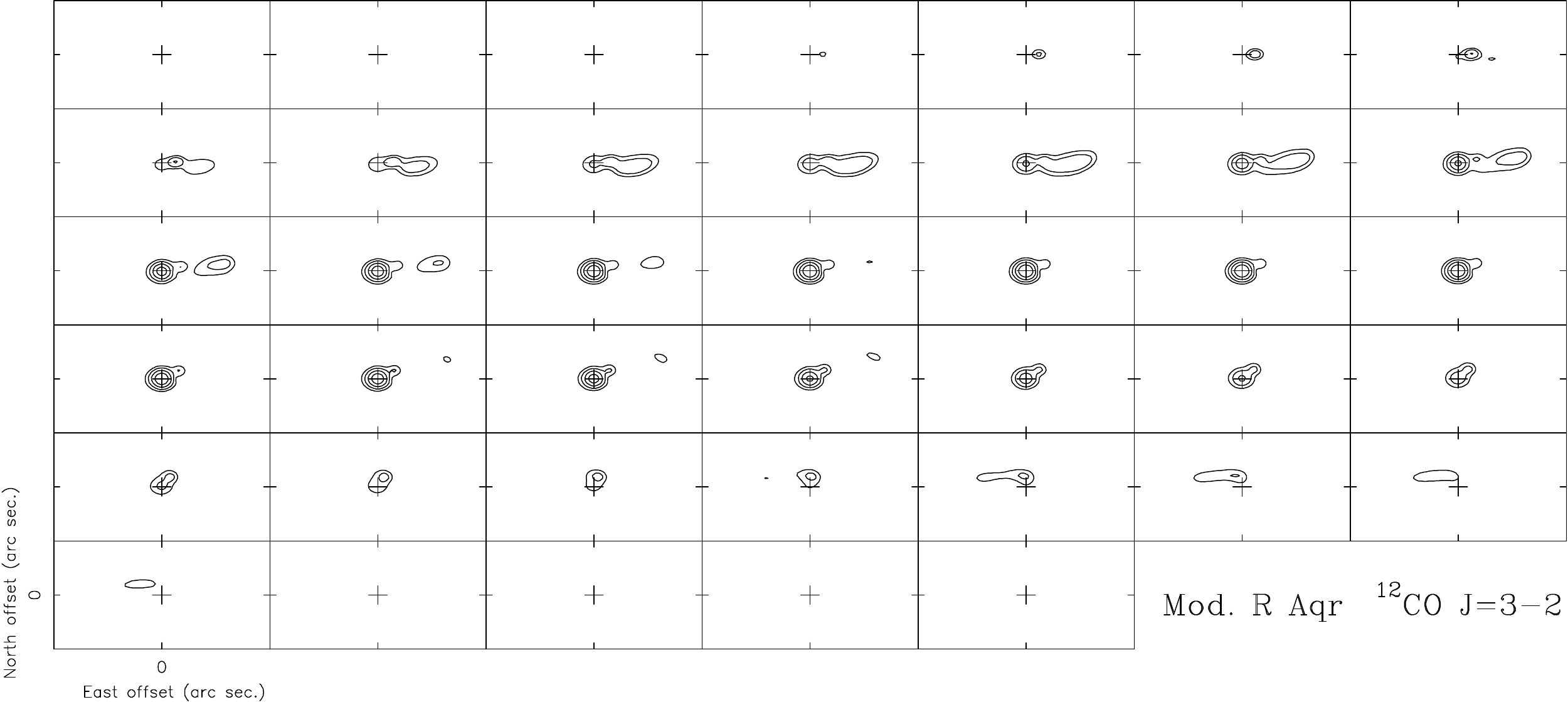}
     \caption{Predictions of our best-fit heuristic model for the
       \doce\ \jtd\ line maps, to be compared with observations,
       Fig.\ \ref{maps32}. All units and scales are the same as in
       Fig.\ \ref{maps32}.}
         \label{pred32}
\end{figure*}

To reproduce the general features of the CO maps (Sect.\ 3), our model
must assume, more specifically, that the emission comes from the inner
circumstellar envelope, whose general structure consists of a {\it
  central condensation}, occupying a region with a size comparable to
the orbit and responsible for the central clump detected in the maps,
and a series of discontinuous {\it spiral arcs} that are strongly focused on
the equatorial plane and are responsible for the plumes detected around the
central clump. The remaining envelope, including the interarms and
regions far from the equator, must contain a very low density of CO
molecules (as indicated by the observations, Sect. 3).

The {\em central condensation} is only poorly resolved by our
mapping. Therefore its structure is not described in detail in the
model and we discuss only global properties, even if this can be very
complex (as shown in Sect.\ 4.2). We just assumed the same kinematics
as in a standard inner AGB shell, with a moderate outward velocity that
increases with the distance. It is remarkable that the observations do
not show any sign of rotation in these inner regions. This is
particularly the case of \trece\ \jtd\ (Fig.\ \ref{maps13co32}), which
comes only from this region and in which no shift in velocity
between the west and east parts of the clump is found.

Hydrodynamical calculations suggest that the {\it spiral arcs} first
include a spiral arm leaving the {\it central condensation},
approximately from the place at which the WD star is expected, and
developing counterclockwise. These CO-rich {\it spiral arcs} in our
model also include the threads of outer spiral(s) that are still
CO-rich. The gas velocity in the arcs must be a composition of
tangential and radial velocities.

The line excitation is described in the code by LTE. This is fully
justified for our low-$J$ transitions (with relatively low
A-coefficients) and under the expected conditions in the inner
circumstellar regions, with high densities \gsim\ 10$^{6}$cm$^{-3}$
(see, e.g., \cite{bujetal10} and the discussion in Appendix A).  With
all these ingredients, we calculated the brightness distribution for
very many lines of sight (i.e., angular offsets with respect to the
nominal center) and velocity shifts (i.e., frequencies corresponding
to Doppler shifts for different LSR velocities) by solving the standard
radiative transfer equation. This distribution is numerically convolved
with the observational beam, yielding maps that are directly comparable
to the ALMA maps (as they show little flux loss). The code is similar
to that we used in previous studies \citep[e.g.,][but we assumed no
  axial symmetry here]{bujetal16}. With this treatment, the
calculations are relatively fast. The full calculation required some
seconds with a PC in standard runs and one or two minutes for the most
accurate calculations.

The model predictions were compared with the ALMA maps. We considered for this purpose the high-quality
\doce\ \jtd\ maps (Fig.\ \ref{maps32}), which show an extended
distribution. We also compared with other CO maps, but less in detail because their quality is lower and they were obtained at different epochs
(as mentioned, Sect.\ 3, comparing data from different epochs in such a
rapidly evolving system is not straightforward).  We show in
Fig.\ \ref{mod} ({\it top}) the density and velocity distributions of
one of the best-fit models in the equatorial plane (i.e., the $z$=0 plane
in our frame). The $y$=0 and $x$=0 planes are also represented in
Fig.\ \ref{mod}. In this best-fit model we also take an angle between
the equatorial plane and the line of sight of 15$^\circ$ and a position
angle of the rotation axis projection on the plane of the sky of 5$^\circ$.  In
Fig. \ref{pred32} we show the \doce\ \jtd\ maps predicted by this
nebula model, to be compared with observational data
(Fig.\ \ref{maps32}; the units and scales are the same in both
images). The comparison is very satisfactory. The
correspondence between the different components in our model and the
predicted features (and corresponding observed features) is very
direct, see the next sub-subsection, showing that there is not much room for
changes in the proposed CO-rich gas distribution and kinematics.

We clearly identify several components in the model. We include
the {\it central condensation}, component {\bf A} in Fig.\ \ref{let},
plus {\it spiral arcs}. These are composed of an inner arc close to
the WD (component {\bf B}), its outer parts (component {\bf C}), an
intermediate small arc ({\bf D}), and a weak outer arc ({\bf
  W}). Other still fainter clumps ({\bf W$'$}) are also considered to
explain some more distant, poorly mapped features. {\bf?}  represents the
strong feature detected at --26 \kms\ LSR, at about 0\secp 1 northeast,
whose origin is not easily understood in terms of the expected
dynamical components. {\bf W$'$} and {\bf ?} are in fact not included in
our modeling (Fig.\ \ref{mod}) and are not discussed in depth.

\subsubsection{Correspondence between the main components of the model
  and the observed features}

\begin{figure}
  
  \hspace{-.3cm}
       \includegraphics[width=9cm]{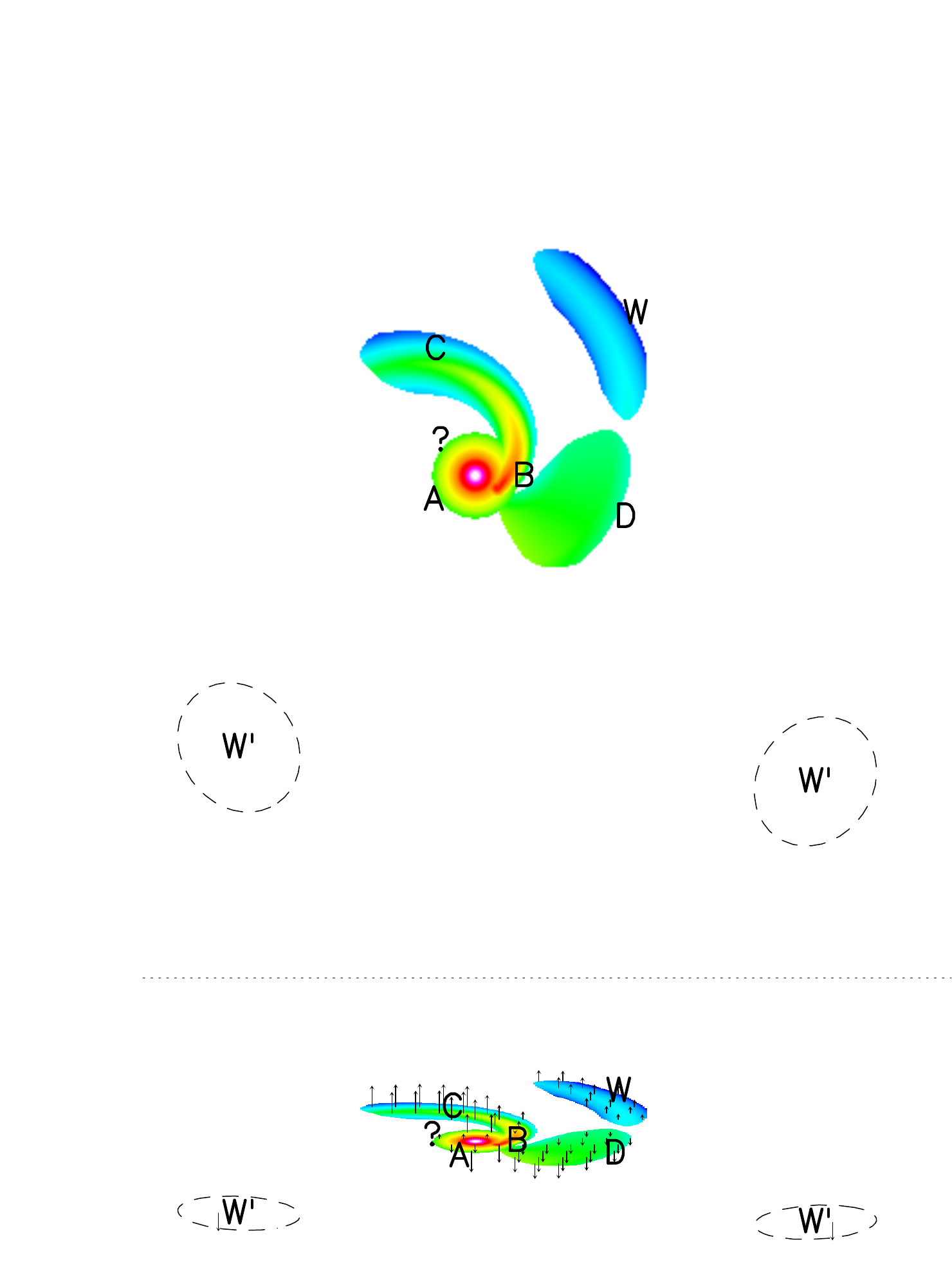}
     \caption{{\it top:} Main components identified in our simple model
       (see Sect. 4.1).  {\it bottom (below the dotted line):}
       Representative image of the equatorial density distribution as
       seen with the actual inclination of the equator. This image is a
       very simplified representation to facilitate interpretation, in
       which the latitude density distribution is not accounted
       for. The projection of the equatorial velocity in the direction
       to the observer is also represented (directly taken from the
       velocity vectors in Fig.\ \ref{mod} {\it top}). Again this is
       only illustrative because the off-equator movements are not
       considered. Downward arrows represent gas approaching the
       observer (relatively negative LSR velocities), while arrows
       pointing upward mean that the gas recedes from us
       (relatively positive LSR velocities).}
         \label{let}
\end{figure}

To better understand the meaning of the various model components, we
also show in Fig. \ref{let} ({\it bottom}) a view of the equatorial density
distribution seen from a certain inclination of the equator with respect to the
line of sight. This representation is only illustrative because, to
avoid confusion, we only accounted for the equatorial density. We also
represent the projection of the gas velocity in the line-of-sight
direction. We again stress that this is just an approximation because
only the equatorial regions and velocities are considered, but we
recall that the density and acceleration due to the WD attraction are
assumed to be high only very close to the plane.

The regions in which the gas recedes from us (vectors pointing upward
in Fig. \ref{let}) must correspond to observed features in the maps with
relatively positive velocities. When the gas approaches us
(vectors pointing downward), the emission must be detected at
relatively negative velocities. Because all the nebula shares
the characteristic circumstellar expansion in some way, regions placed behind the
star should tend to emit at positive velocities. The rotation
introduced by the gravitational interaction should yield negative
velocities for nebular regions placed close to the WD (which is placed
slightly rightward in our representation and approached us
during the observing run). We mentioned that both expansion and
rotation are accounted for in our model.

\begin{figure}
     \begin{center}
       \includegraphics[width=8.5cm]{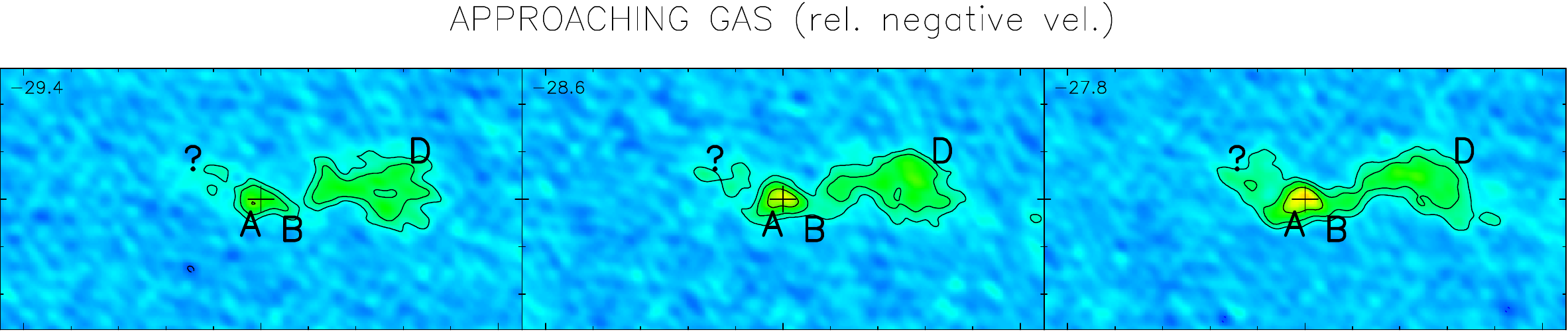}
       
       \vspace{0.3cm}
       \includegraphics[width=8.5cm]{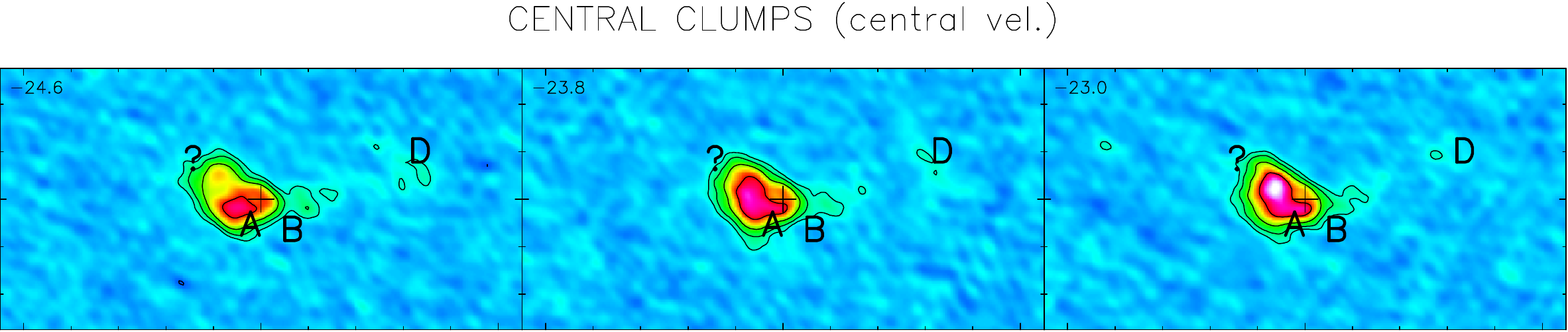}
       
              \vspace{0.3cm}
       \includegraphics[width=8.5cm]{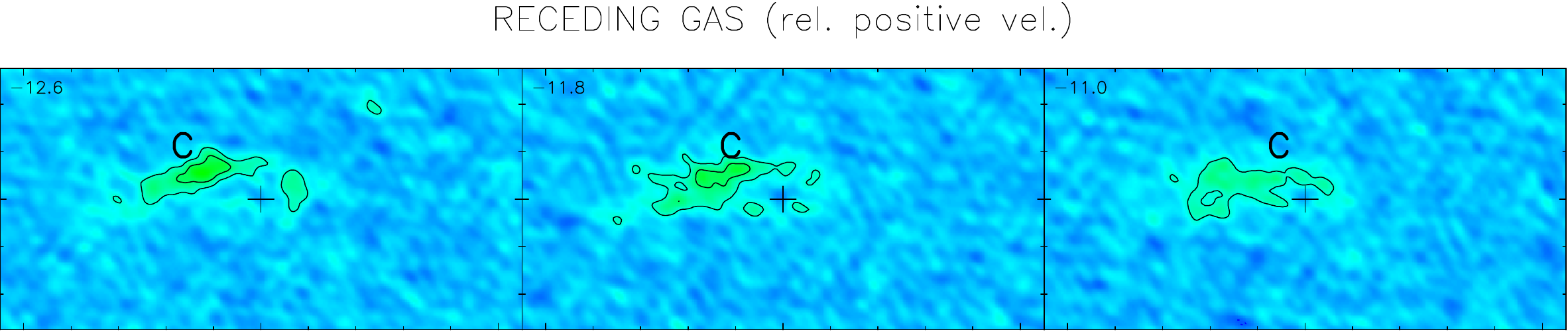}
              
       \vspace{0.3cm}
       \includegraphics[width=8.5cm]{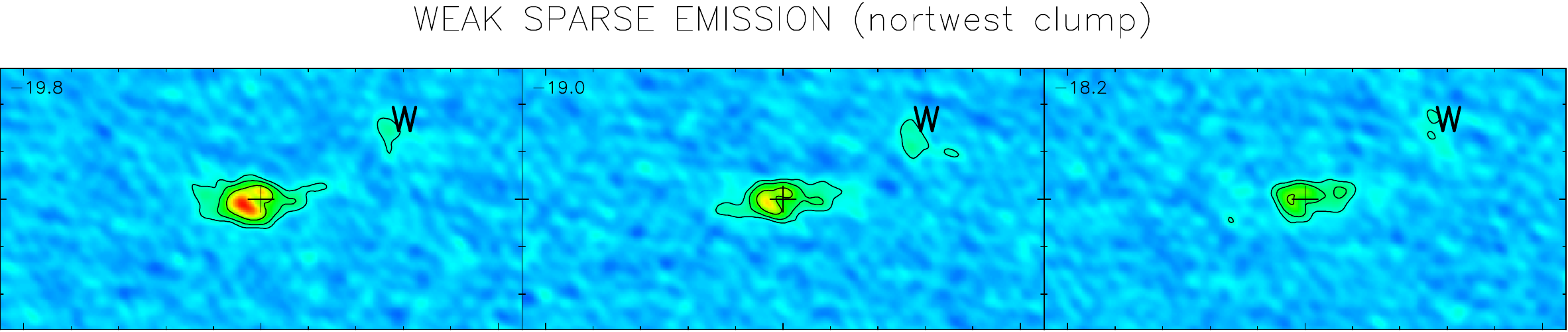}
       
       \vspace{0.1cm}
       \caption{Representation of the different observed features
         (from Fig.\ \ref{maps32}) and the components of our model
         ({\bf A}, {\bf B}, {\bf C}, {\bf D}, {\bf W,} and {\bf ?} in
         Fig.\ \ref{let}) that are responsible for the
         corresponding predicted emission.  }
       
         \label{ident}
     \end{center}
     
\end{figure}

In Fig.\ \ref{ident} we show maps for representative velocity
channels, in which the correspondence between the observed features and
the above components is indicated. The central condensation must show
both positive and negative (moderate) projected velocities as a result
of the expansion previous to interaction with the WD. Component {\bf A}
in Fig.\ \ref{let} can therefore explain the observed central
condensation with a moderate velocity dispersion around the central one
(feature {\bf A} in Fig.\ \ref{ident}); note that the meaning of {\it
  central velocity} in our case is not obvious, the velocity of the
system barycenter is $\sim$ $-$24.5 \kms\ LSR \citep{alcmik21}, but the
AGB star showed a slightly more positive velocity during the
observations. In component {\bf B}, the expansion projection is almost
negligible and the rotation is expected to be high. These components
explain the negative-velocity feature ({\bf B} in Fig.\ \ref{ident}).
It is possible that both expansion and
rotation contribute to explain the negative velocity in {\bf D}. In the outer
regions of the spiral arm, {\bf C}, the projected velocity is dominated
by the expansion, and it is responsible for the arc that is clearly detected at
relatively positive velocities. Finally, we introduced diffuse
outer arcs {\bf W, W$'$}. {\bf W} can explain the sparse very weak
clumps detected in the outer regions at northwest offsets. {\bf W$'$}
vaguely represents two of the weak southern clumps detected in the
low-resolution maps (Fig.\ \ref{ext32}). These components probably
are remains of the very outer regions of the {\bf B-C} arm (see a tentative
general view in Fig.\ \ref{modfot}). {\bf W$'$} was not
included in our modeling because their nature was too uncertain. We
recall again that these weak components may show a clumpy
emission pattern that is not clearly defined in our data.

In our maps, we also identify a conspicuous feature that cannot fit the
general features of our modeling, see label {\bf ?} in Figs. \ref{let},
\ref{ident}. If it is equatorial, this component should be placed behind the
star, but it shows a negative velocity projection, which cannot be
explained by our model (with just expansion and rotation
velocities). We suggest that it is the result of the emission of a
CO-rich gas clump falling back to the AGB star and is perhaps not exactly
placed on the equator, which would remain CO-rich because of shielding
by the AGB star surroundings. Inflows due to complex hydrodynamical
interactions are sometimes predicted by model calculations, mostly in
off-equator regions (see Sect.\ 4.2), but the description of their very
complex structure and dynamics is beyond the scope of our simple
modeling, and we only point out its existence.

We stress that some quantitative results in our descriptive modeling
are necessary to explain the observations, provided that the general
assumptions (our ``axioms'' described above) are true. We list those
results below.

\noindent $\bullet$ The general structure of the CO-rich gas in our
model, with the central condensation and several arcs placed close to
the equator, is necessary provided that we assume equatorial symmetry
and that we know with reasonable accuracy the inclinations with respect
to the north and to the line of sight. The observed features directly
correspond to those of the model, and more extended CO-rich gas is
improbable.

\noindent $\bullet$ The sizes of the model components are also
necessary, within the (very good) angular resolution of our
observations, scaled for the value of the distance (which gives the
conversion from angular to linear units), and depending on other
assumptions. For instance, the size of the central clump, component
{\bf A}, must be of about 7 10$^{14}$ cm; the uncertainty, $\sim$ 2
10$^{14}$ cm, due to the resolution and to the contribution of nearby
components, is not negligible. The contribution of other components is
also the main source of uncertainty for the shape and extent of {\bf
  B}; they must be accurate if the identification scheme depicted in
Fig.\ \ref{ident} is correct. The outermost arcs ({\bf D} and {\bf W})
must reach a distance $\sim$ 1.5 10$^{15}$ cm, which is an accurate
estimate: the most distant emission in Fig.\ \ref{maps32} reaches
$\sim$ 0\secp 4 and, because of its low declination offset, these
regions must be placed roughly perpendicular to the line of sight. The
outer part of the first arm (component {\bf C}) must be placed at
$\sim$ 7 10$^{14}$ cm. This value depends on the adopted inclination
with respect to the line of sight. For an inclination of 30$^\circ$,
the distance to {\bf C} would be twice smaller, and {\bf C} would be
1.5 times farther for inclinations of about 10 degrees. Out of this
range (already generously wide), {\bf C} could not be considered as a
continuation of {\bf B}.

\begin{figure*}[h] 
  \includegraphics[width=\textwidth]{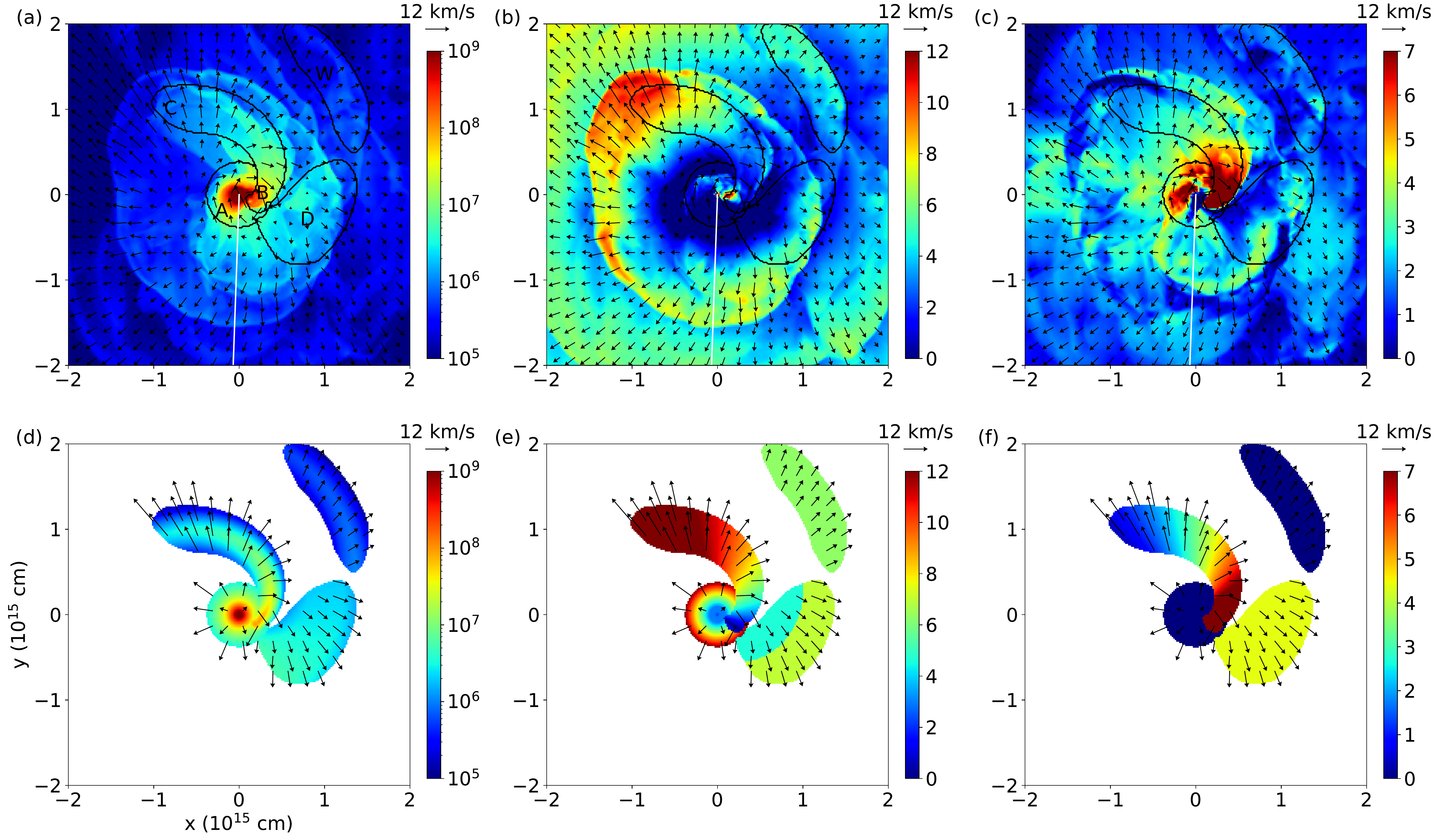}
  \caption{\label{fig:cmp}
    {\it top}: Hydrodynamical model versus ({\it bottom}) our heuristic
    model for (a, d) number density, (b, e) radial velocity, and (c, f)
    rotational velocity distributions of the gas in the equatorial plane.
    Color bars label in units of cm$^{-3}$ for number density and \kms\
    for velocities. The overlaid arrows represent the gas velocity vectors.
    The pericenter of WD is located on the straight white line, and the
    current orbital phase of the binary stars is at 65$^\circ$ before the
    pericenter. 
  }
\end{figure*}

\begin{figure}[h] 
  \center
  \includegraphics[width=0.5\textwidth]{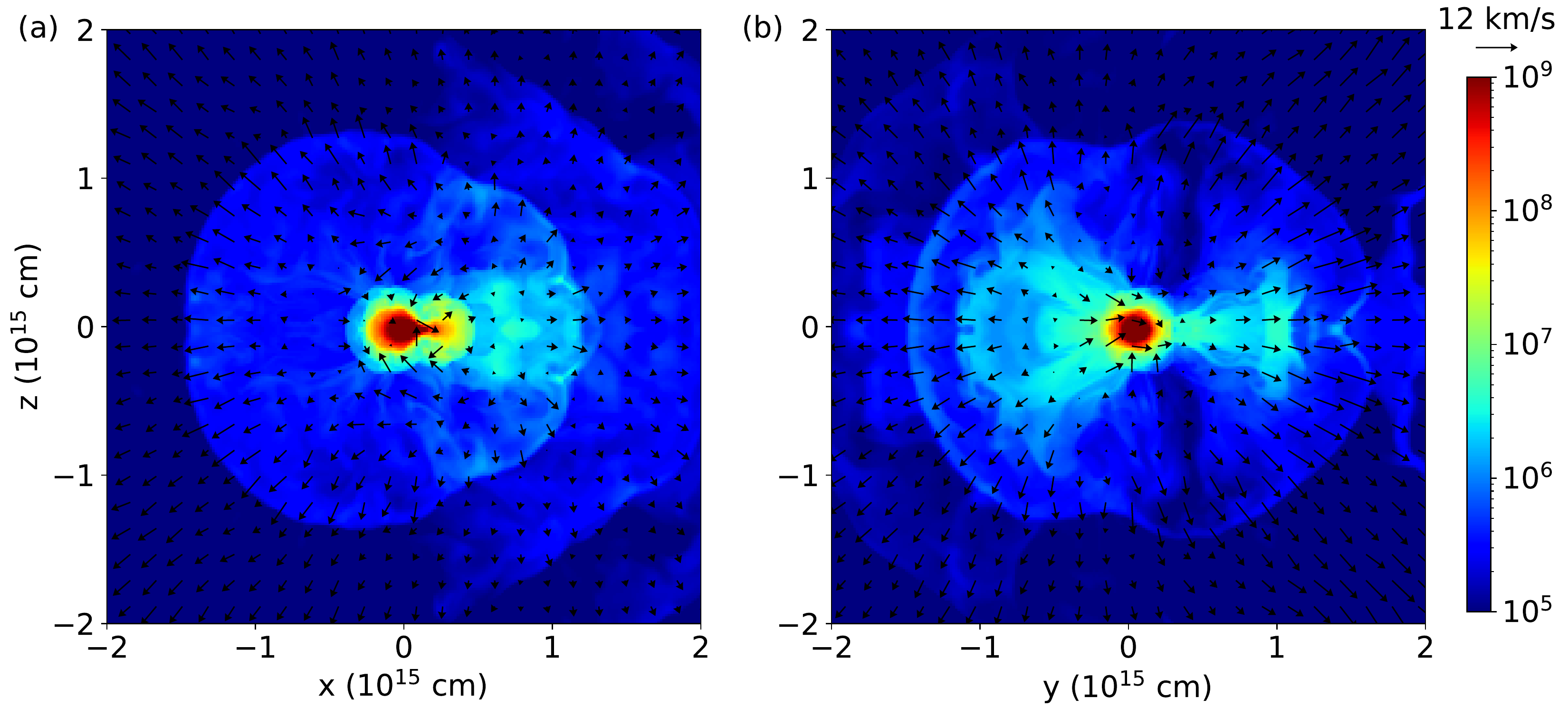}
  \caption{\label{fig:den} Density distribution of the gas in the
    hydrodynamical model in the (a) $y$=0  and (b) $x$=0 planes,
    revealing the vertical flattening onto the equatorial plane. The
    velocity vectors show the fallback of gas toward the equatorial
    plane and the AGB star from up to the height of 50 AU.  }
\end{figure}

\noindent $\bullet$ Some properties of the velocity field in our model
are also necessary to describe the observations. The expansion velocity
of the central clump is directly given by the dispersion of the central
intense feature in the maps, within 1 \kms. The expansion velocity of
component {\bf C} is accurately given by the shift in velocity of the
corresponding observed feature, 10-15 \kms; however, its 
  rotational
velocity, which points more or less perpendicularly to the line of
sight, is only poorly determined. In contrast, the rotational
velocities of components {\bf B} and {\bf D}, which basically point to
us, are accurately determined and must be of 3-10 \kms\ to
explain the observed blueshifts.  Expansion velocities in {\bf W} of
about 5-9 \kms\ (with low rotation) are also necessary, although in
this case, the weak detected emission and the high angles between the
velocity and line-of-sight vectors may significantly affect our
estimates.  All these velocities depend slightly on the inclination of
the equator with respect to the line of sight (provided that it is
\lsim\ 30$^\circ$) because all components are basically placed in the
equator.  It is important to note that some of the kinematical items of
our description show a certain degree of degeneracy. Rotational and
radial velocities could change in some areas, notably {\bf B} and {\bf
  D}, provided that the projected velocity in the direction of the
observer is kept the same. Component {\bf D} could be slightly shifted
to positive values of the y-axis in Fig.\ \ref{mod}, partially
occupying the region between {\bf B} and {\bf W}; then we would have to
introduce a higher rotation velocity that is hardly compatible with
the hydrodynamical calculations in Sect.\ 4.2 (although we would achieve a better
fit of the observed maps). In these cases, we generally tried to
adapt our heuristic model to the results obtained from hydrodynamical
calculations.

\subsection{Hydrodynamical simulations of the interaction
  between the compact companion and the AGB wind. Comparison with our
  heuristic model}

We performed hydrodynamical simulations with the stellar and
orbital parameters of R Aqr binary stars using the code FLASH 4.5
\citep{fry00}. The fluid motion is governed by the continuity,
momentum, and energy equations for hydrodynamics, including the
gravitational attraction toward the individual stars, and the
intrinsic acceleration of AGB wind mimicking radiation pressure onto
dust particles \citep[for details, see, e.g.,][]{kim13, kim19}. We also
implemented radiative cooling of the gas and accretion sink onto the
white dwarf (Lee et al., in prep.).

\begin{figure*}[h] 
  \includegraphics[width=\textwidth]{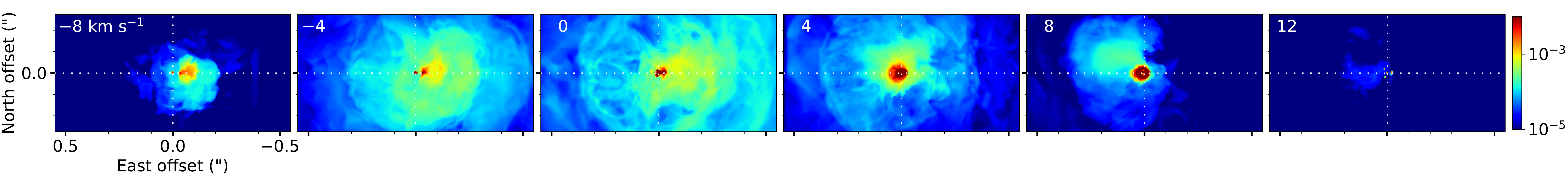}
  \includegraphics[width=\textwidth]{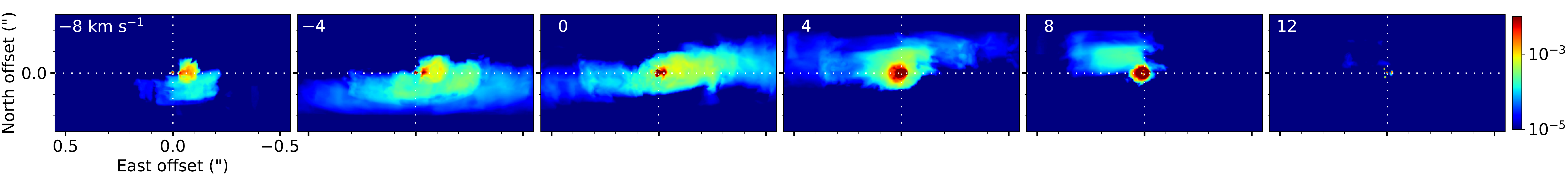}
  \caption{\label{fig:chm}
    {\it top}: Column-density channel maps of our hydrodynamical model at
    an inclination of 15$^\circ$ from the edge-on view of the equatorial plane,
    and ({\it bottom}) the same, but with a pseudo-restriction of the density
    distribution up to a height of 15 AU.
    The central velocity of each channel with respect to the systemic
    velocity ($\sim$ $-$24.5 \kms\ LSR)
    is indicated at the top left corner of each panel, and the individual
    channel width is 4\,\kms.
    Color bars are in units of $\rm g\,cm^{-2}\,(\kms)^{-1}$.
  }
\end{figure*}

\begin{figure}[h] 
  \center
  \includegraphics[width=0.5\textwidth]{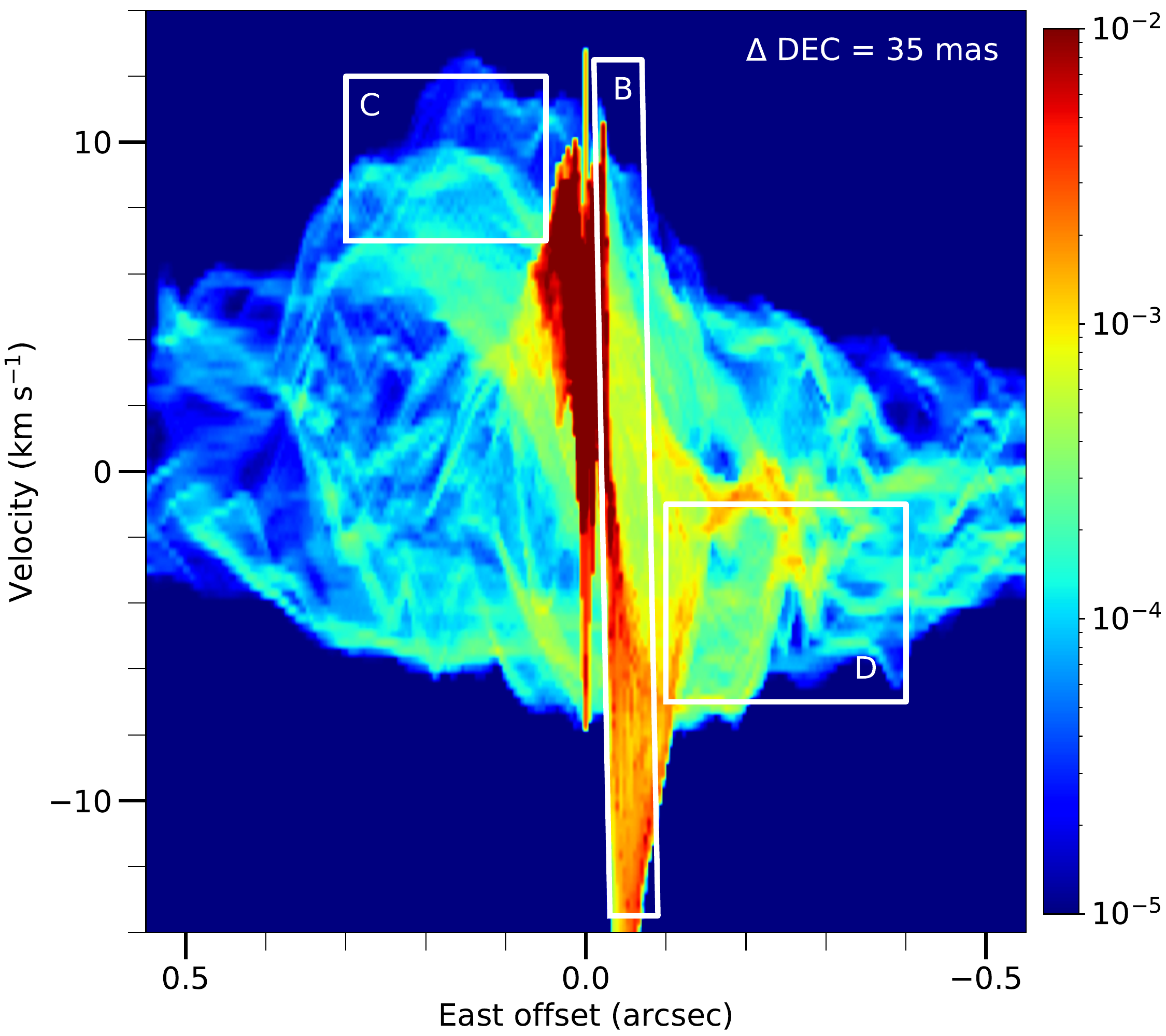}
  \caption{\label{fig:pvd}
    Position-velocity diagram of our hydrodynamical model at an 
    inclination of 15$^\circ$ from the edge-on view of the equator.
    The velocity is measured with respect to the systemic velocity. 
    The color bar is in units of $\rm g\,cm^{-2}\,(\kms)^{-1}$. See
    observations in
     \cite{bujetal18}.
  }
\end{figure}

In the simulations, we adopted the AGB and WD masses of 1.41\,\ms\ and
0.82\,\ms, respectively, and the average orbital separation (sum of the
semimajor axes) of 15.7\,AU, yielding an orbital period of 42 yr. The
mass-loss rate of the AGB star was adopted to be 3 10$^{-6}\,\ms\,\rm
yr^{-1}$. An orbital eccentricity of 0.57 was applied, causing the
orbital separation to vary from 24.6 AU at the apocenter down to 6.8 AU
at the pericenter. The orbital velocity of the WD varied from
3.7\,\kms\ at the apocenter to 13.5\,\kms\ at the pericenter, which is
comparable to the expansion velocity of the AGB wind. The sufficiently
fast orbital velocity relative to the radial wind velocity causes the
``transverse'' velocity component in the flow near the WD near its
pericenter, while the flow tends to pass the WD straight at its
apocenter. As a consequence, the flow near the stars becomes very
complex over the entire orbital phase. Although some hydrodynamic
simulations in the literature treated highly eccentric orbit cases such
as $e=0.8$ \citep[e.g.,][]{kim17,kim19,randall20}, their lagging
stellar orbital velocities made the situation relatively simpler than
the model visualized here.

We iterated hydrodynamic simulations to examine the wind velocity
acceleration with the fixed stellar and orbital parameters. The wind
velocity near the stars (in particular, the velocity near the WD)
controls the formation of the accretion disk, the portion of wind escaping
from the gravitational potential of stars, and the overall shape of
the spiral (see details in Lee et al., in prep.). The situation
becomes more complicated when the orbital shape is noncircular: the
spiral pattern is then pushed out faster at the pericenter and
relatively more slowly at the apocenter.

Our basic nebula description (Sect. 4.1) is quite similar to some of
the hydrodynamical predictions, see Fig.\,\ref{fig:cmp}, but in our
descriptive model only some parts of the spiral structure appear and
the contrast between them and the surrounding regions is stronger. We
attribute these differences to the selective photodissociation of CO by
the WD UV field, see the discussion in Sect. 4.3.

The main features of the hydrodynamical model exhibited in
Fig.\,\ref{fig:cmp} are listed below.

\noindent $\bullet$
For the {\bf A}, {\bf B}, {\bf C}, {\bf D}, and {\bf W} components
defined in Sect. 4.1, their locations and density levels match the
dense structures in the equatorial plane in the outcome of the
hydrodynamical simulation reasonably well. The significant drop in
spiral density following the C region is of particular interest. It is
consistent with the ALMA images showing a relatively small extension
toward the east.

\noindent $\bullet$ Component {\bf B},  indicating the spiral arm at high
density attached to the WD, is clearly identified in our hydrodynamical
model. At the head of the spiral in this model, the flow mostly shows
a rotational motion in the clockwise direction with the velocity of
11-13\,\kms\ on average within the scale height of $\sim$ 5 AU. In
the immediate neighborhood of the WD, the rotational velocity of the
flow increases to 28\,\kms.

\noindent $\bullet$ The spiral tail following the WD is well located
within the {\bf C} region defined in the heuristic model. The expansion
velocity of this spiral pattern is up to 11\,\kms, which is slightly
lower than the velocity criteria in Sect. 4.1. However, the
three-dimensional data cube of the hydrodynamical model shows that, 
above the equatorial plane, the spiral pattern shifts toward the $+y$
direction, which would be degenerate in the observation of the R Aqr
binary system as the orbit is viewed close to but not exactly
edge-on. The expansion velocity of this spiral-shell pattern approaches
 12\,\kms, with some variations along its vertical stretch
from the equatorial plane (see Fig.\,\ref{fig:den}b).

\noindent $\bullet$ The abrupt density decrease beyond the {\bf B}-{\bf
  C} spiral arm, which appears in both the
observations and the hydrodynamical calculations, is
probably due to the high eccentricity of the orbit. A circular-orbit
simulation, which was tested with the same orbital parameters except
for the orbital shape, does not show this density discontinuity. In
general, it exhibits  smooth undulations that extend in a highly isotropic
manner, which contradicts the observed pattern.

\noindent $\bullet$ The {\bf D} region shows the convergence and
divergence of many different branches of density flows caused by
various velocity components due to the initial AGB wind acceleration
process, the effects from eccentric orbital motions of the stars, and
some fallback flows toward the equatorial plane. The net expansion and
rotational velocities of the flows within region {\bf D}  are not higher
than 6.5 and 4.5\,\kms, respectively, in agreement with our
descriptive model.

\noindent $\bullet$ The arc appearing in region {\bf W}  has a
shell structure with an expansion velocity up to
6.5\,\kms\ considering a typical height of 15 AU. The rotational
velocity is low and does not exceed 2\,\kms.

\noindent $\bullet$ Finally, we stress the presence of some infall
toward the AGB star (Fig.\ \ref{fig:den}; notably from the other side
of the WD). As discussed in Sect.\ 4.1, this feature may qualitatively
explain the observed component {\bf ?}, which is otherwise very difficult to
understand, provided that some of the infalling gas remains CO-rich.

In order to mimic the observing view for the nebulae around the R Aqr
binary system, independently of the comparison with the heuristic
modeling, we rotated the hydrodynamical model cube about the $x$-axis
until the equatorial plane is at 15$^\circ$ from the line of sight. The
WD approached us and was located in the southern part.
Fig.\,\ref{fig:chm} shows the column-density channel maps of our
hydrodynamical model at six representative velocities. In general, the
channels near the systemic velocity show superfluous features compared
to the ALMA maps per velocity channel, which would be mostly invisible
in CO line transitions due to photodissociation (see
Sect. 4.3.). Nevertheless, we can obtain some useful information from
this simulation by comparing with the ALMA channel maps, in particular,
for the relatively less-structured channels at high velocities. We
found that the expansion velocity of component {\bf C} is up to
10\,\kms\ (with respect to the systemic velocity) in the equatorial
plane, but increases to 12\,\kms\ at its vertical stretch in the
off-plane. This component is apparent in the channel map at 12
\kms\ with respect to the systemic velocity as a nearly horizontal
feature to the northeast. It corresponds to the emission in the ALMA
channel maps at the {\it LSR} velocities of $-14.2$ to
$-10.2$\,\kms\ (see Fig. 1). The column-density channel map at
$-8\,\kms$ shows the enhanced density distribution to the southwest
0\farcs2, while the corresponding ALMA map ({\it LSR} velocities near
$-31.8\,\kms$) is extended mostly to the west up to 0\farcs3. We also
have an indication for the {\bf ?} component from the north-northeast
stretch in the $-4\,\kms$ column-density map (top panel). Its
disappearance in the bottom panel might further imply a survival of
this dense gas cloud at some height from the equator against the
photodissociation effect described in Sect. 4.3.  Finally, in
Fig.\,\ref{fig:pvd}, we show a position-velocity diagram of column
density along the east-west central cut with the width of the slice
corresponding to the beam size of $^{12}$CO 3--2 maps. Fig.\ 13
displays the asymmetric velocity distribution in the redshift to the
east up to $\sim+11\,\kms$ (component {\bf C}) compared to the
blueshift up to $\sim-7\,\kms$ ({\bf D}). The very large velocity
dispersion of component {\bf B} is also clearly visible.

\subsection{Effects of photodissociation on the extent and shape
  of the CO-rich region}

Except for H$_2$, CO is the most abundant molecule in circumstellar
envelopes around AGB stars and shows the largest extent
because it is very efficiently selfshielded against interstellar UV
radiation \citep{mamon87,cernicharo15}. A strong
emitter of UV radiation in the hot and dense inner regions of the
envelope, as it is the case of the white dwarf in the R\,Aqr symbiotic
stellar system, might induce important changes on the
distribution of CO, compared with standard AGB envelopes. The first
important effect is related to the morphology of the whole gaseous and
dusty envelope, as a consequence of the gravitational interaction with
the white dwarf (see Sects.\ 3, 4.2). The second important effect is
related to the photodissociation induced by UV photons emitted by the
white dwarf, which can affect the distribution of molecules, including
CO, within the envelope. 

\begin{figure}
     \begin{center}
       \includegraphics[width=9cm]{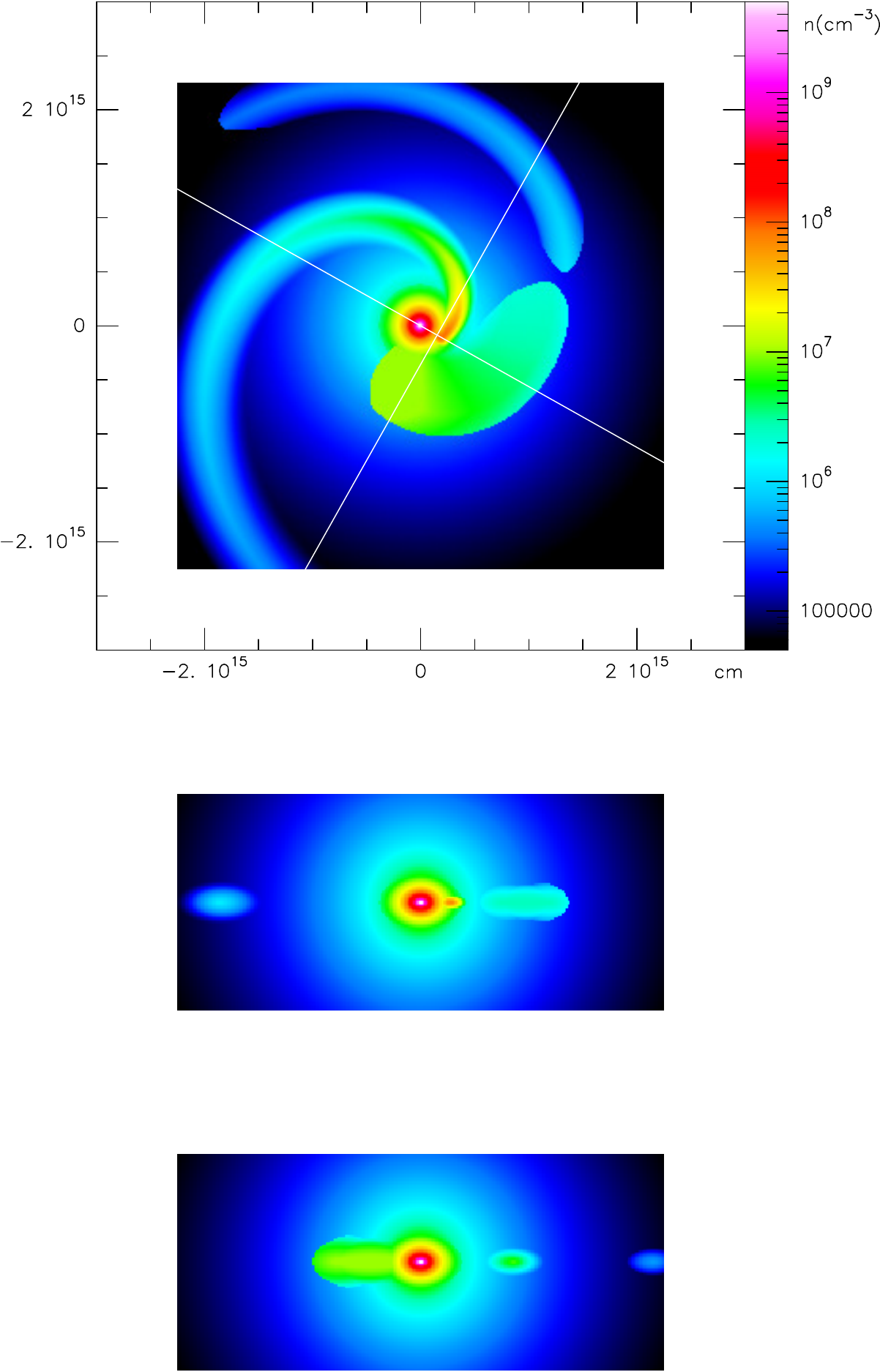}
       \vspace{.1cm}
     \caption{Equatorial density distribution of the gas in the extended
       version of our best-fit model, which intends to include not
       only CO-rich gas, but also the adjacent regions in which
       molecules are assumed to be photodissociated.  (See Sect.\ 4.3
       for details.)  {\it top:} Density distribution in the plane of
       the equator. {\it middle:} Density distribution in the y=0
       plane. {\it bottom:} Same, but for the x=0 plane. All
       representations share the same units and scales as in
       Fig.\ \ref{mod}.}
         \label{modfot}
     \end{center}
\end{figure}

As mentioned (Sect.\ 1), the properties of compact companion are
difficult to determine.  In addition, we must realize that the derived
parameters mostly refer to the emission of the WD and its surroundings,
including the probable accretion disk. The radiation parameters of the
UV source were discussed by \cite{kaler81}, \cite{burgarella92},
\cite{meierk95}, and \cite{schmid17}. Their estimates of the
temperature range from $\approx$ 40 to 80 kK, and for the luminosity
from $\approx$ 0.05 \ls\ to 5-20 \ls.  For the purpose of this study,
we therefore decided to make our own estimates. We used two
long-exposure IUE spectra of the central source of R Aqr taken on 14
July 1991 (SWP 42069 and SWP 42070). The coadded spectra revealed a
rising continuum for $\lambda \la 1700\, \AA,$ while the equivalent
width of HeII\,1640 emission line, $EW(1640)=9.9\pm1.3\,\AA$, is
consistent with an ionizing source with and effective temperature $T
\sim 70000$ K. In this analysis, we assumed a typical reddening E(B-V)
$\sim$ 0.23 mag along the line of sight toward the WD, which is
somewhat lower than the average values found toward the Mira
(Sect.\ 1), according to results usually found in SSs
\citep[see][]{mikol99, gromadzki09}. We note that a significantly
higher extinction toward the WD is also incompatible with the
appearance of the IUE spectrum. Assuming a blackbody distribution of
the hot WD continuum, we fit the R-J tail to selected continuum windows
shortward of $\lambda \sim 1700\, \AA$. The best fit to the UV
continuum and HeII\,1640 emission line flux was obtained for $T \sim
80$ kK and $R_{WD}/d \sim 1.1 10^{-12}$, which, assuming a distance of
265 pc, gives $L \sim 6$ \ls. The radius of the hot WD is consistent
with a CO WD radius with a mass of $\sim$ 0.7-0.8 \ms\ (compatible with
our orbital solution). Similarly, we derived $T \sim 70$ kK and $L \sim
4.5$ \ls\ from the coadded IUE spectra taken in 1980-1986.
Accordingly, we finally adopt $T=80000$ K and $L=5$ \ls\ as
representative values to be used for analyzing the effects on molecular
photodissociation.

In regard to photodissociation, we expect that the UV photons emitted
by the white dwarf first create a relatively compact HII region
around it, where UV photons with energies higher than 13.6 eV are
fully absorbed, and a second region located farther away, showing the
typical stratification C$^+$/C/CO of dense photon-dominated regions
(PDRs; \citealt{tielens85}). That is, we could expect a more or less
spherical region void of CO around the white dwarf. Depending on
various key parameters, such as the mass-loss rate of the AGB star,
the separation between AGB and WD, and the effective temperature and
luminosity of the WD, the effect on the CO distribution can be very 
different. It ranges from a small CO-depleted region around the white
dwarf, which may be difficult to probe through observations, to a
strong effect on the CO envelope, leading to complex morphologies or
even to a complete removal of the CO envelope in certain SSs.

Modeling the photodissociation of CO in such a system becomes
challenging because of several complications. One of them is that the
spherical shape of the envelope can be strongly distorted by the
gravitational effect of the WD, leading to complex spiral-like
structures and a complicated kinematics in which expansion and
rotation are mixed. The relative abundance of dust is highly uncertain in the
surroundings of the AGB star because dust is expected to form at a
certain distance from the red giant. Moreover, the orbital timescales can
be comparable to the (photo)chemical timescales, creating a
time-dependent problem in which the shape of the CO-rich envelope can
vary with time.

When we take the high prevalence of binary stellar systems into
account, interest in modeling the effect of internal UV sources in AGB
envelopes is timely (e.g., \citealt{saberi19}), although for the
moment, models cannot solve the complications described above. We
evaluated the effects of UV radiation from the WD on the distribution
of CO in the R\,Aqr system by using the Meudon PDR code
\citep{lepetit06}, in which we implemented a correction to take the
geometrical dilution of the UV field (and of photodissociation rates)
with increasing distance from the WD into account. We evaluated the
chemical composition at steady state along four representative
directions from the white dwarf lying in the orbital plane (at
longitudes $\Lambda$ = 0$^{\circ}$, 90$^{\circ}$, 180$^{\circ}$, and
270$^{\circ}$, where $\Lambda$ = 0$^{\circ}$ corresponds to the
direction toward the AGB star and the angles increase counterclockwise;
see Fig.\ \ref{modfot}). We adopted an extended version of our
heuristic model (Sect.\ 4.1), in which we crudely extended the
description of the density distribution also to regions that,
presumably because of photodissociation of carbon monoxide, are in fact
not observed in CO lines. We therefore recall that, because of the lack
of empirical information and the complex structure suggested in the
hydrodynamical models, this extension is very tentative. We also
explored models in which we decreased the density by factors of two and
three because the determination of the column densities of gas and dust
from our line emission fitting is relatively uncertain: it depends
significantly on various parameters, such as the fractional abundance
of CO, the conversion from gas to dust densities, and the actual
velocity dispersion. We adopted solar elemental abundances and the
gas-phase chemical network used by \cite{agundez18} to model
protoplanetary disks, which also host dense, warm, and UV-illuminated
regions, as in R\,Aqr. Optical and infrared observations of Mira-type
(O-rich) AGB stars have shown that the most relevant elements (C, N,
and O) have essentially solar abundances
\citep[e.g.,][]{smith86,galan17}.  Elemental abundances have not been
determined in symbiotic systems with a Mira star, such as R Aqr, but we
can reasonably expect that elemental abundances would not deviate much
from solar. Moderate deviations would not affect the main conclusions
obtained from the PDR models, although they may slightly hinder the
comparison between total column densities in our heuristic model and in
chemical models.

We considered dust extinction typical of the interstellar medium,
but with a dust-to-gas mass ratio half of the interstellar value to
account for the probable depletion of dust in the surroundings of the
AGB star, where dust has not yet fully formed. The kinematics of the
envelope is complex, and this might reduce the selfshielding
of CO. Because the Meudon PDR code does not allow treating velocity
gradients, we adopted a relatively high turbulent velocity of 15
km s$^{-1}$ to reduce the optical depth of the UV lines of CO and the
extent of selfshielding.

\begin{figure*}[ht!]
\begin{center}
\includegraphics[width=0.323\textwidth]{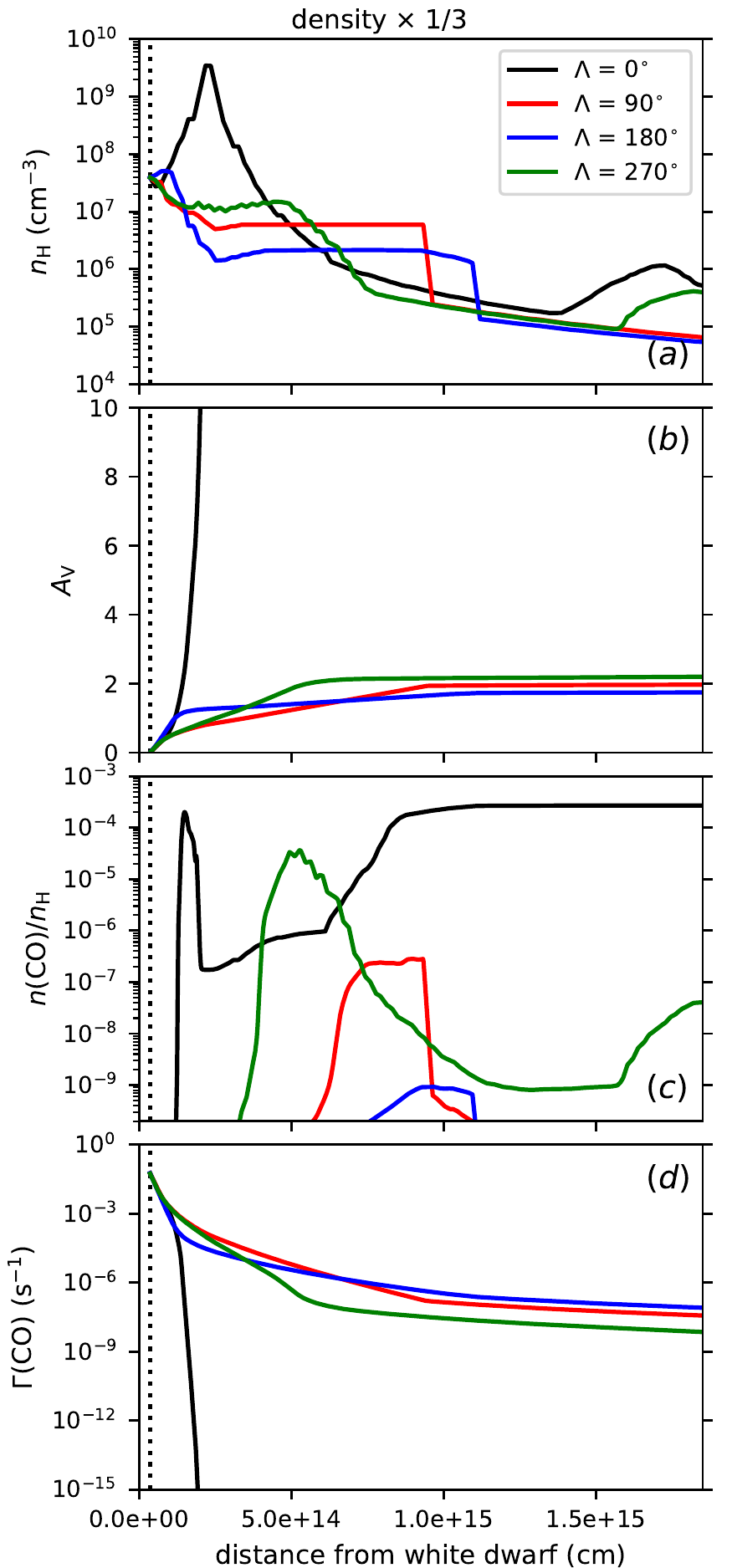} \hspace{0.1cm}
\includegraphics[width=0.323\textwidth]{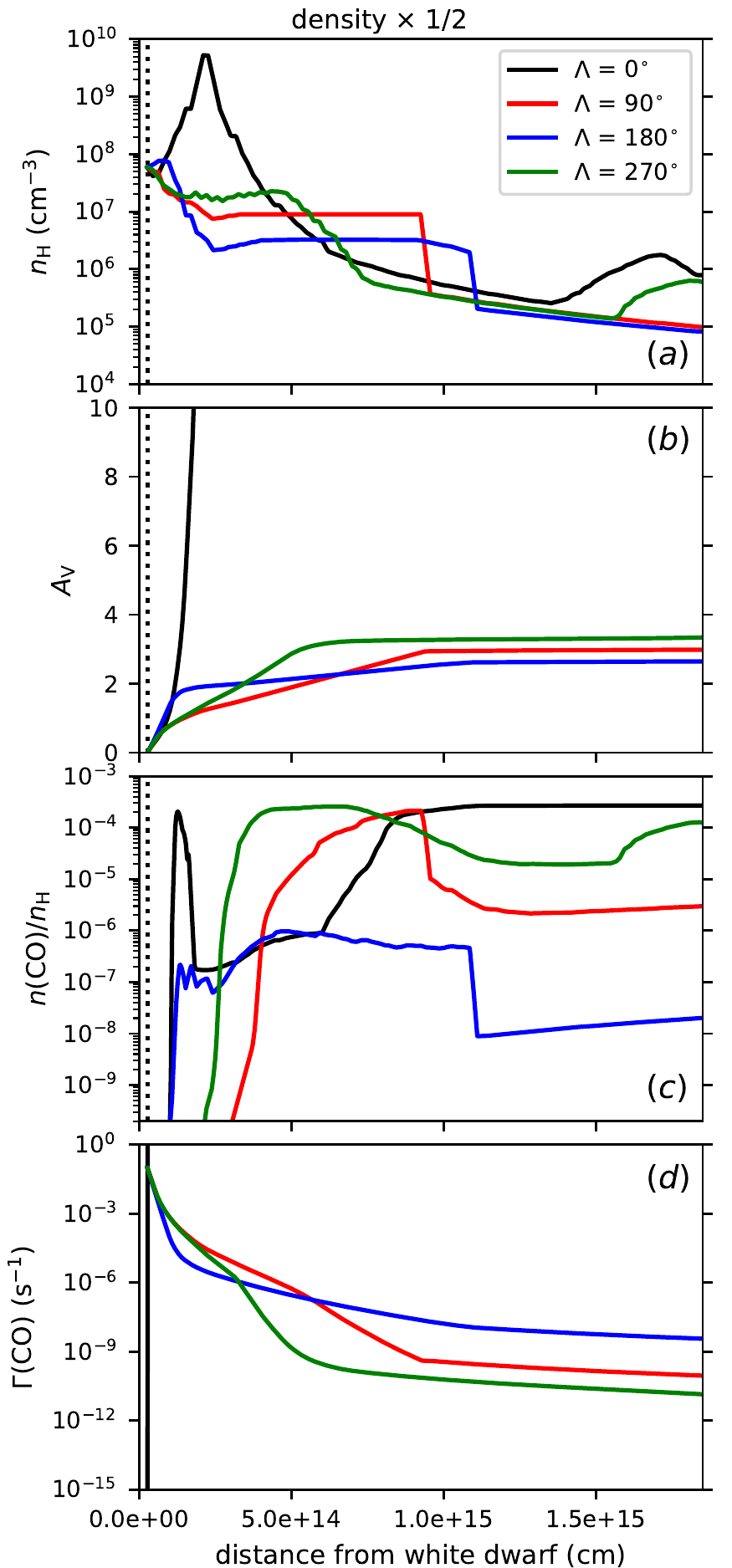} \hspace{0.1cm}
\includegraphics[width=0.323\textwidth]{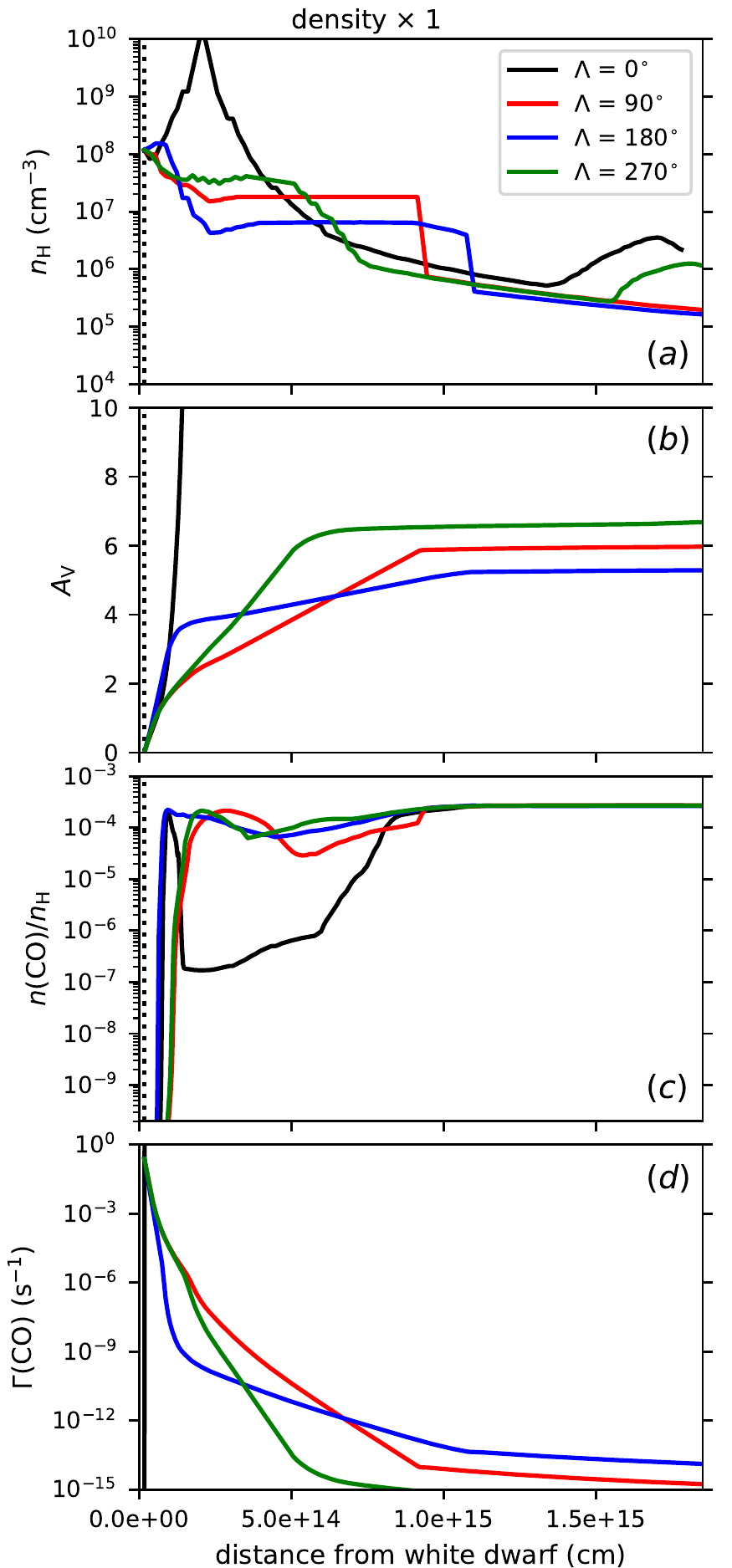}
\vspace{.1cm}
\caption{Selected input and output quantities of
    the PDR models shown as a function of the distance from the
    white dwarf. We consider three models, shown from left to right,
    in which the density is one-third, half, and one times that
    of the extended heuristic model (see text), respectively. The different panels
    show from top to bottom $(a)$ the density of H nuclei $n_{\rm
      H}$ (note the peak reached at a distance of $\sim$ 2 10$^{14}$
    cm for $\Lambda$ = 0$^{\circ}$, which stands for the position of
    the AGB star), $(b)$ the visual extinction $A_{\rm V}$ measured
    from the WD,  $(c)$, the fractional abundance of CO relative to H
    nuclei $n$(CO)/$N_{\rm H}$, and $(d)$ the photodissociation rate
    of CO, $\Gamma$(CO). The vertical dotted line at the left edge is
    the position of the Str\"omgren radius.}
\label{fig:pdr}
\end{center}
\end{figure*}

For the conditions adopted here to model the photodissociation of CO in
R\,Aqr, we find that the C$^+$/CO transition lies at a visual
extinction of about 2 mag from the WD. In the extended heuristic model
(panels on the right in Fig.~\ref{fig:pdr}), the adopted densities
result in total visual extinctions higher than $\sim$ 5 mag for any
direction in the orbital plane, implying that the C$^+$/CO transition
lies close to the WD and that beyond this point, CO is very abundant
regardless of the longitude (see panel $b$ on the right in
Fig.~\ref{fig:pdr}). In the direction toward the AGB star, at $\Lambda$
= 0$^{\circ}$, CO is depleted at distances between 1.5\,10$^{14}$ cm
and $8\,10^{14}$ cm (see panel $c$ on the right in Fig.~\ref{fig:pdr}),
however, because CH$_4$ becomes the main C-bearing molecule. This fact
is predicted by the chemical model for a very concrete region of R\,Aqr
(inner shells on the other side of the WD) as a consequence of the lack
of UV and high temperatures prevailing there (500-1000 K). It has been
also predicted to occur in dense, warm, and UV-illuminated regions of
protoplanetary disks \citep{agundez08}. It is difficult to verify a
phenomenon like this in our case because of the complex CO-emission
structure, but we do not find any clear sign of strong CO depletion in
the close surroundings of the primary (CO is also found to be abundant
in the inner shells around standard AGB stars). This feature arises in
the model because chemical reactions tend to incorporate carbon into
CH$_4$ rather than into CO at high temperatures. In R Aqr, however, CO
is expected to be formed in the close surroundings of the AGB star in
thermochemical equilibrium conditions, and, if sufficiently shielded
from the UV radiation, the possibility is low that it is dissociated to
liberate atomic carbon and to allow chemical reactions to convert it into
CH$_4$ in times shorter than the system evolution time. We therefore
consider that the displacement of CO by CH$_4$ is unlikely to occur in
wide regions around R Aqr.

If the density of our extended heuristic model is reduced by factors
of two and three, then the total visual extinctions in the equatorial
plane lie in the range 2-4 mag (except at $\Lambda$ = 0$^{\circ}$,
where the total $A_{\rm V}$ is much higher because of the high density
close to the AGB; see panels $b$ in the middle and on the left in
Fig.~\ref{fig:pdr}). When we take into account that the visual extinction
measured to the AGB star in R\,Aqr from the Earth is $\sim$ 1-2 mag
(Sect.\ 1), these models with the reduced density are likely to
reproduce the conditions prevailing in R\,Aqr better. We recall
that the measured values of the extinction, see Sect.\ 1, correspond to
the line of sight, which is not in the equatorial plane, and they must
be lower than the extinction values in Fig.\ \ref{fig:pdr}.
Interestingly, when the density is progressively reduced by factors of
two and three, CO starts to be severely depleted along the direction at
$\Lambda$ = 180$^{\circ}$ (see panels $c$ in
Fig.~\ref{fig:pdr}). This direction indeed has the lowest total
$A_{\rm V}$ (see panels $b$ in Fig.~\ref{fig:pdr}), leading to a higher
photodissociation rate for CO (see panels $d$ in Fig.~\ref{fig:pdr})
and a lower abundance for this molecule. The directions at $\Lambda$ =
90$^{\circ}$ and 270$^{\circ}$ have intermediate values of total
$A_{\rm V}$ and densities of the correct order to allow for a competition
between the formation of CO through chemical reactions and destruction through
photodissociation, resulting in variable fractional abundances. On the
other hand, the direction toward the AGB star, at $\Lambda$ =
0$^{\circ}$, has very high densities and CO is efficiently protected
against UV photons from the WD. At directions beyond the
orbital plane, the total visual extinction falls below 2 mag, resulting
in high photodissociation rates and low CO abundances. As
the density decreases, CO therefore starts to differentiate severely depending
on the direction from the WD in the orbital plane, while it vanishes
far from the equator. For example, in the model in which the
density is reduced by a factor of two, CO has a high abundance at
$\Lambda$ = 0$^{\circ}$, variable abundances at $\Lambda$ =
90$^{\circ}$ and 270$^{\circ}$, a low abundance at $\Lambda$ =
180$^{\circ}$ (see panel $c$ in Fig.~\ref{fig:pdr}, middle column), and
a negligible abundance out of the orbital plane (not shown in
Fig.~\ref{fig:pdr}).  This behavior is consistent with our observations
(Sect.\ 4.1).  In the model in which the density is reduced by a factor
of three, CO is strongly dissociated, except in the direction toward
the AGB star (see panel $c$ on the left in
Fig.~\ref{fig:pdr}).  We have focused our discussion on
  the case of \doce\ because other molecules, for which selfshielding
  is much weaker, are much more easily dissociated and expected to show
  very compact extents, in agreement with observations \citep[see
    Sects.\ 2 and 3 and further discussion in][]{gomezg21}.

It is also worth discussing the time-dependent nature of the R\,Aqr
system and the timescales of the various competing processes. The
white dwarf takes $\sim$ 42 yr to complete an orbit. A first question
to address is whether CO molecules are photodissociated sufficiently
rapidly compared to the orbital evolution of the system. The
photodissociation rate depends on the position in the envelope and the
$A_{\rm V}$ toward the WD. For directions with moderately low visual
extinctions ($A_{\rm V} \sim$ 2), the photodissociation rate of CO is
about 10$^{-8}$ s$^{-1}$ at 2\,10$^{15}$ cm (see panels $b$
and $d$ in Fig.~\ref{fig:pdr}), which corresponds to $\sim$ 3 yr and
thus occurs much faster than the orbit of the WD. At shorter radial
distances, CO molecules will be photodissociated even faster. Only for
directions with high visual extinctions from the WD, CO
photodissociation will take longer than the orbital period. A second
relevant question is whether in the nebular regions that are more exposed to
the UV photons from the WD, where CO molecules have been efficiently
photodissociated, there is sufficient time to replenish CO before the
WD returns and illuminates them again with UV radiation. From
pseudo-time-dependent gas-phase chemical models (see, e.g.,
\citealt{agundez13}) it was found that the formation timescale of CO
through chemical reactions, $\tau_{\rm f}$, is inversely proportional
to the density of H nuclei $n_{\rm H}$ through the expression $\tau_{\rm
  f}$ = $f$/$n_{\rm H}$, where the scaling factor $f$ is $\sim$
2\,10$^9$ yr cm$^{-3}$ to form CO with its maximum abundance and $\sim$
10$^7$ yr cm$^{-3}$ to form it with about 10\% of the maximum
abundance. It follows that for a density $n_{\rm H}$ = 10$^5$ cm$^{-3}$
(reached at 2\,10$^{15}$ cm; see panel $a$ in the middle and on the
left in Fig.~\ref{fig:pdr}), the time to form 10\% of the maximum
abundance of CO is 100 yr. It is therefore likely that chemical
reactions cannot fully replenish CO during an orbit of the WD. On the
other hand, the time needed to inject fresh CO molecules from the AGB
wind, assuming it has not already been photodissociated by the WD, at
distances of 2 10$^{15}$ cm is about 60 yr (assuming an expansion
velocity of 10 km s$^{-1}$), which is somewhat longer than the orbital
period. That is, there is room for some replenishment of CO, although
as soon as the UV field from the WD illuminates these
regions sufficiently, CO will be rapidly photodissociated. In particular, this
affects the regions that are located at distances larger than 2\,10$^{15}$ cm,
which are therefore expected to be void of CO.

In summary, the main lesson from these models is that, for visual
extinctions of a few mag, there is room for significant changes in the
CO chemistry between different directions from the white dwarf, with a
difference in the abundance of orders of magnitude. Moreover, if
densities and photodissociation rates result in comparable formation
and destruction rates, the abundance of CO can show interesting
modulations along a given direction, with significantly lower
abundances in interarms. These features are observed in our maps of CO
and are roughly reproduced by the models, which are likely to describe
the conditions in R\,Aqr reasonably well because they result in visual
extinctions similar to those that are observed. The particular
conditions of R\,Aqr imply that after CO is photodissociated in the
regions that are more exposed to the UV radiation from the white dwarf,
it is unlikely that CO can be replenished rapidly enough compared to
the orbital time of the system. A more general conclusion is that, for
values of $A_{\rm V}$ that are significantly lower than a few mag, CO
is expected to be largely photodissociated, and we therefore expect a
CO-poor envelope. On the other hand, if $A_{\rm V}$ is significantly
higher than a few mag, CO should be efficiently shielded against
photodissociation and a large CO-rich envelope is expected. This
general conclusions probably apply to other binary systems that are
composed of an AGB star and a white dwarf.

\section{Conclusions}

We have performed ALMA observations of R Aqr, a well-known SS. This
system is composed of an AGB star plus a hot and compact WD, orbiting
with a relatively long period of about 42 yr. The system is known to
show strong interaction and spectacular symbiotic phenomena. The stars
are surrounded by an extended nebula that is detected at several
wavelengths, which shows an equatorial structure plus axial
high-velocity jets, probably the result of accretion by the compact
stellar component of material previously ejected by the AGB star. In
Sect.\ 1 we discussed the properties of this system that are relevant
for our study. Molecular lines are typically very weak in SSs; R Aqr is
by far the symbiotic system that shows the richest and strongest
molecular emission, although it presents very weak CO lines compared
with standard AGB stars (Sect.\ 3).

We present high-quality maps of R Aqr in several CO lines, namely
\doce\ \jdu, \jtd, and \jsc\ and \trece\ \jtd\ (Sects. 2 and 3). For
\doce\ \jdu\ and \jtd, maps using different array configurations have
been obtained, leading to higher and lower angular resolutions. The
corresponding results have been compared, also taking single-dish data
into account, and we deduce that most CO emission is detected in the
interferometric maps, which are a good representation of the actual
CO-emitting region.  The analysis of the maps of different lines
indicates that the \doce\ emission is partially opaque and comes from
gas with a relative high excitation, typically with several hundred K.

The observations show a strongly structured and compact emitting
region, occupying less than 1\secp 5. CO is emitted from a central
intense clump, about 0\secp 2 wide, plus an elongated region formed by
various arcs, which look like shreds of spiral arms. From a comparison
with our knowledge on the binary system, we deduce that the relatively
extended arcs are significantly focused on the equator of the
system. The brightness distributions observed in R Aqr are highly
different from those of standard AGB stars, in which observations
similar to ours usually show arcs or incomplete spiral arms, but within
an extended overall structure that is spherical or shows a moderate
axial symmetry (Sect. 3).

Spiral structures in the nebula around SSs are predicted by
hydrodynamical models of gravitational interaction between the wind of
the primary and the compact companion.  We have developed a heuristic,
descriptive model of the CO-emission maps (Sect.\ 4.1) based on
general results from hydrodynamical modeling. The model is 3D,
 and the geometry only assumes plane symmetry with
respect to the equator. From the given distributions of the density,
temperature, and velocity of the CO-rich gas, our code 
predicts velocity-channel maps that can be directly compared with the
observations.

Our best-fit descriptive model, which is able to reproduce the
observations quite satisfactorily, is composed of a central expanding
condensation and several arcs showing both expansion and rotation
movements. We identified a number of components (Sect.\ 4.1) in the
CO-rich nebula: a central condensation (component {\bf A}), a first arm
associated with the WD (components {\bf B} and {\bf C}), and outer arms
(components {\bf D} and {\bf W}, which are sometimes weak and
patchy). There are also still weaker clumps ({\bf W$'$}), as well as a component
of uncertain origin (labeled {\bf ?}) that could be due to infalling
material.  The central condensation occupies about 5 10$^{14}$ cm,
slightly more than the orbit, and the arcs occupy about 2 10$^{15}$ cm
in the equatorial plane, significantly less in altitude from the
equator. The involved velocities range between 3 and 15 \kms. The
fastest rotation appears close to the WD.
Our modeling confirms that only some threads of the spiral arms are
detected, the remainder appears to show no significant CO emission.

As mentioned, this heuristic nebula model is compatible with our
general ideas on hydrodynamical interaction in an SS. We also performed
sophisticated calculations specifically for this case, see
Sect. 4.2. This hydrodynamical modeling includes the gravitational
attraction of both stars on the circumstellar material, as well as the
hydrodynamical basic physics and radiative cooling, leading to nebular
structures that depend on the orbital phase. Although our comparison
of the two models can only be semiquantitative, the agreement between
the hydrodynamical predictions and our heuristic/descriptive model of
the CO line emission is very satisfactory, provided that we also take
into account the high degree of CO photodissociation expected in some
regions (see below). All the different components of our heuristic
modeling can also be identified in the hydrodynamical simulations
(Sect.\ 4.2, Fig.\ \ref{fig:cmp}). We underline 
that our simulations assuming a relatively high eccentricity show
prominent {\bf C} and {\bf D} components, notably with a density cavity
in between, in agreement with the well-defined observed features, a
property that does not appear in calculations assuming a circular
orbit. The model also predicts infall in certain regions.

Even with the caveat that photodissociation significantly affects the
distribution of CO molecules, several discrepancies subsist. The
central condensation shows a very complex dynamics in the
hydrodynamical modeling that is dominated by rotation and also shows
expansion and infall. We did not try to incorporate this intricacy into
our simple heuristic modeling (but assumed an isolated-AGB-like
kinematics in the central region); in addition, the predicted rotation
would yield a variation in velocity along the east-west direction that
is not observed. The structure and dynamics of the spiral arms are
remarkably well compatible with the hydrodynamical calculations,
including the rotation and expansion velocity fields, except that the
velocity moduli tend to be slightly higher in the descriptive models
(i.e., in the observations).

Hydrodynamical models in general yield nebulae that are significantly
more extended than observed in R Aqr. These predictions also show
density contrasts that are in general much lower than those
suggested by our ALMA maps and their 3D description. We interpret these
differences as due to photodissociation by the UV radiation from the
hot WD, which is compatible with the lack of molecular emission in SSs
(Sect.\ 4.3). We have performed a detailed analysis of the chemistry of
the nebula around R Aqr. We discussed photodissociation by UV from the
WD and selfshielding effects, CO reformation within a very complete
chemical reaction network, and time-dependent phenomena due to the
orbital movement. In general, we found that when the UV radiation is not
strongly shielded, photodissociation is very fast and CO is
destroyed. CO formation, which mainly depends on the density, is often
slower.  We conclude for R Aqr that this object shows a set of relevant
parameters such that nebular CO is not photodissociated in some areas,
but completely disappears in others, namely in regions with lower
densities and  that are more directly exposed to the WD radiation. (Chemical
calculations tend to require slightly lower densities than those found
from our heuristic model, but they are still compatible within the
uncertainties.)  In general, we expect a significant lack of CO in the
interarms, regions far from the equator, and in equatorial regions at
distances \gsim\ 2 10$^{15}$ cm.  Because all these regions are
cyclically exposed to the UV following the orbital periods, the time
dependence of the chemistry also helps to explain the lack of CO,
particularly in the outer regions. These results from our chemical
modeling are compatible with our description of the maps of R Aqr, at
least qualitatively. They also explain the lack of molecular emission
in other SSs, most of which show shorter orbital periods and thinner
nebulae.

\begin{acknowledgements}
  We are grateful  to the anonymous referee for his/her thorough
  reading of the paper and to Dr.\ S.\ Mohamed for enlightening discussions
  during the first phases of this project.  This work has been supported
  by the Spanish MICINN, grants AYA2016-78994-P, AYA2016-75066-C2-1-P,
  PID2019-105203GB-C2, PID2019-106110GB-I00, and PID2019-107115GB-C21,
  and by the National Science Centre, Poland, grant OPUS
  2017/27/B/ST9/01940. M.A.\ acknowledges funding support from the
  Ram\'on y Cajal programme of Spanish MICIU (grant RyC-2014-16277) and
  from the European Research Council (ERC Grant 610256:
  NANOCOSMOS). H.K.\ acknowledges support by the National Research
  Foundation of Korea (NRF), grant funded by the Korea government
  (MIST) (No. 2021R1A2C1008928). This paper makes use of the following
  ALMA data: ADS/JAO.ALMA\#2017.1.00363.S and 2018.1.00638.S. ALMA is a
  partnership of ESO (representing its member more states), NSF (USA)
  and NINS (Japan), together with NRC (Canada), MOST and ASIAA
  (Taiwan), and KASI (Republic of Korea), in cooperation with the
  Republic of Chile. The Joint ALMA Observatory is operated by ESO,
  AUI/NRAO and NAOJ.
\end{acknowledgements}

\appendix

\section{Limits of our LTE treatment of the CO excitation}

In this appendix, we estimate the error we can make by assuming
that the CO levels are in thermal equilibrium at the local
temperature. We performed calculations of the populations by solving
the statistical equilibrium equations under the assumption of very low
optical depths. We showed that the CO rotational lines are mostly
optically thin. The advantage of using this approximation is that the
level populations do not depend on the various approximations we can
follow to perform the calculations (LVG-like local approximations or
nonlocal treatments for different gas distributions). The calculations
were performed for the relevant physical conditions, that is, for typical
values of the density and temperature in relevant regions of the model
nebula, because other properties of the nebula have no effect under the
optically thin approximation. We discuss  the effects of
this and other implicit assumptions below, but the comparison
is very conclusive in any case.

\begin{figure}

     \begin{center}  
     \hspace{1cm}
       \includegraphics[width=9cm]{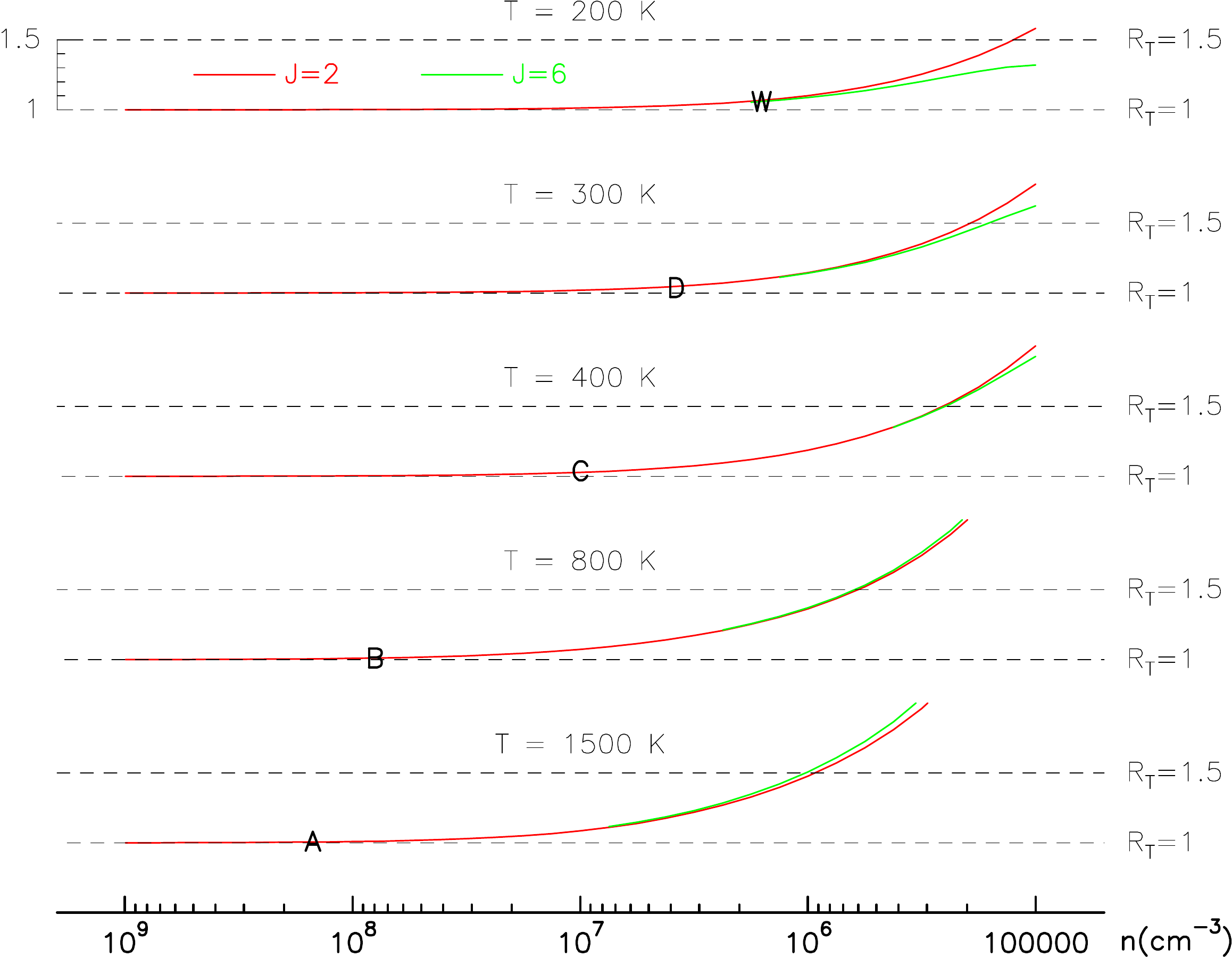}
       \vspace{0cm}
     \caption{Ratios of the relative level
       populations calculated accurately by solving the statistical
       equilibrium equations for CO, assuming very low optical depths,
       and those calculated assuming thermalized level
       populations, $R_{\rm T}$. See the discussion in the text. We
       calculated $R_{\rm T}$ for the $J$=2 and $J$=6 levels and for
       the conditions expected in our main model features {\bf A},
       {\bf B}, {\bf C}, {\bf D}, and {\bf W}, as indicated in the
       figure. The deviations from thermal equilibrium populations are
       in all cases $<$ 10\%.}
         \label{rt}
     \end{center}
\end{figure}

In Fig.\ \ref{rt} we show the ratios, $R_{\rm T}$, of the
relative level populations calculated by solving the statistical
equilibrium equations and the thermalized populations. We show the
population ratios for the $J$=2 and $J$=6 levels; intermediate levels
(like $J$=3) show intermediate behaviors.  The variations
in intensities of the lines that leave or arrive at these levels
must be of the same order as the variations in the level populations.
The typical conditions of the main components considered in our
modeling are represented in the figure. The deviation
from thermalized populations is always negligible, smaller than 10\% 
and very similar for the different levels, which results in negligible
effects in the synthetic maps.  Departures always tend to yield
higher populations and therefore (slightly) higher
intensities. The reason is that for the relatively high involved
temperatures, far higher than the relevant level energies, the
effect of underthermalization is that levels with very high
$J$-numbers are less densely populated than in thermal equilibrium; the
excitation temperatures of low-$J$ levels are much less affected
because the relevant transitions, such as those we observe, have much
smaller A-coefficients. The result is that molecules must crowd
together in these lower levels and the transitions joining them are
(in our case, slightly) more intense that in the thermalized case.

These estimates are simple, but our approximations are not expected to
affect our conclusions. We list the expected effects below.

\noindent $\bullet$
To simplify our discussion in this appendix, we performed statistical
equilibrium calculations assuming optically thin lines. Some opacity
might be present, particularly if the nebulae shows clumpiness at a very
small scale. However,  opacity always tends to increase
the thermalization degree (except for a few pathological cases,
certainly not that of the single-ladder level structure of
CO). Accounting for a certain degree of opacity will therefore lead to
still more thermalized level populations and values of $R_{\rm T}$
still closer to 1. We performed LVG calculations for optically
thick cases and the other physical conditions adopted here, and this
result was always confirmed: our calculations for low optical depths
represent upper limits to the actual departures from thermalized
populations.

\noindent $\bullet$ We have claimed that CO is underabundant in less
dense regions (e.g., far from the equator), and for this reason, they
are not detected. The values of $R_{\rm T}$ clearly tend to increase
for lower densities. However, this effect would lead to relatively more
intense emissions from these regions, which would lead to a still
stronger discrepancy between predictions and observations. An ``exact''
treatment of the excitation in low-density regions will therefore
reinforce our conclusion that CO is strongly underabundant in them.

\noindent $\bullet$ Other uncertainties might be present, for instance, in the
collisional transition coefficients. However, our results
mostly depend on the level-integrated inelastic collisions, which have been well
known for many years. Moreover, the departures form thermal
equilibrium are very small in our case, and the effects of the uncertainties on
collisional cross sections are very small, far smaller than the
various sources of uncertainty in our general discussion. This result
is also widely confirmed by various calculations: to increase $R_{\rm
  T}$ by more that 1\%, we must change the cross sections by more than
25\%.

\end{document}